\documentclass{aa}

\usepackage[varg]{txfonts}
\usepackage{natbib}
\usepackage{blindtext}
\usepackage{booktabs} 
\usepackage{pifont}
\usepackage{lscape}
\usepackage{multirow}
\usepackage{wasysym}
\usepackage{cleveref} 
\usepackage{upgreek}

\usepackage{color}

\newcommand{\WA}[1]{\ensuremath{\mathrm{WA}_{#1}}\xspace}

\newcommand{\nh}{\ensuremath{N_\text{H}}\xspace}
\newcommand{\logxi}{\ensuremath{\log \xi}\xspace}
\newcommand{\swift}{\textit{Swift}\xspace}
\newcommand{\xmm}{\textit{XMM-Newton}\xspace}
\newcommand{\rxte}{\textit{RXTE}\xspace}
\newcommand{\suzaku}{\textit{Suzaku}\xspace}
\newcommand{\spitzer}{\textit{Spitzer}\xspace}
\newcommand{\hst}{\textit{HST}\xspace}

\begin{document}

\title{A Variable-Density Absorption Event in NGC 3227 mapped with \textit{Suzaku} and \textit{Swift}}

\author{T.~Beuchert\inst{\ref{inst1},\ref{inst2}}
\and A.G.~Markowitz\inst{\ref{inst1},\ref{inst3},\ref{inst4}}
\and F.~Krau\ss\inst{\ref{inst1},\ref{inst2}}
\and G.~Miniutti\inst{\ref{inst5}}
\and A.L.~Longinotti\inst{\ref{inst6}}
\and M.~Guainazzi\inst{\ref{inst7}}
\and I.~de La Calle P\'erez\inst{\ref{inst7}}
\and M.~Malkan\inst{\ref{inst8}}
\and M.~Elvis\inst{\ref{inst9}}
\and T.~Miyaji\inst{\ref{inst10},\ref{inst3}}
\and D.~Hiriart\inst{\ref{inst10}}
\and J.M.~L\'{o}pez\inst{\ref{inst10}}
\and I.~Agudo\inst{\ref{inst11}}
\and T.~Dauser\inst{\ref{inst1}}
\and J.~Garcia\inst{\ref{inst9}}
\and A.~Kreikenbohm\inst{\ref{inst1},\ref{inst2}}
\and M.~Kadler\inst{\ref{inst2}}
\and J.~Wilms\inst{\ref{inst1}}
}

\institute{Dr.\ Remeis-Sternwarte \& Erlangen Centre for Astroparticle
  Physics, Universit\"at Erlangen-N\"urnberg, Sternwartstrasse 7,
  96049 Bamberg, Germany 
\label{inst1}
\and Lehrstuhl f\"ur Astronomie, Universit\"at W\"urzburg, 
Emil-Fischer-Straße 31, 97074, W\"urzburg, Germany
\label{inst2}
\and Center for Astrophysics and Space Sciences, University of
California, San Diego,  9500 Gilman Dr., La Jolla, CA 92093-0424, USA
\label{inst3}
\and Alexander von Humboldt Fellow
\label{inst4}
\and Centro de Astrobiolog\'ia (CSIC–INTA), Dep. de Astrof\'isica,
European Space Astronomy Centre, PO Box 78,  Villanueva de la
Ca\~nada, 28691 Madrid, Spain 
\label{inst5}
\and Instituto de Astronom\'ia, Universidad Nacional Aut\'{o}noma de
M\'{e}xico (UNAM), 04510 Ciudad de M\'{e}xico, D.F. M\'{e}xico
\label{inst6}
\and European Space
Agency, European Space Astronomy Centre, PO Box 78, Villanueva de la
Ca\~nada, 28691 Madrid, Spain
\label{inst7}
\and Physics and Astronomy Department, UCLA, Los Angeles, CA 90095-1562, USA
\label{inst8}
\and Harvard-Smithsonian Center for Astrophysics, 
60 Garden St., Cambridge, MA 02138, USA
\label{inst9}
\and Instituto de Astronom\'{i}a, Universidad Nacional Aut\'{o}noma de
M\'{e}xico, Km 103, Carret, Tijuana-Ensenada, Ensenada 22860, BC,
Mexico (P.O.\ Box 439027, San Ysidro, CA, 92143, USA)
\label{inst10}
\and Instituto de Astrof\'{\i}sica de Andaluc\'{\i}a (CSIC), Apartado
3004, E-18080 Granada, Spain 
\label{inst11}
} \date{xx-xx-xx / xx-xx-xx }

\abstract {The morphology of the circumnuclear gas accreting onto
  supermassive black holes in Seyfert galaxies remains a topic of much
  debate. As the innermost regions of Active Galactic Nuclei (AGN) are
  spatially unresolved, X-ray spectroscopy, and in particular
  line-of-sight absorption variability, is a key diagnostic to map out
  the distribution of gas.}{Observations of variable X-ray absorption
  in multiple Seyferts and over a wide range of timescales indicate
  the presence of clumps/clouds of gas within the circumnuclear
  material. Eclipse events by clumps transiting the line of sight
  allow us to explore the properties of the clumps over a wide range
  of radial distances from the optical/UV Broad Line Region (BLR) to
  beyond the dust sublimation radius. Time-resolved absorption events
  have been extremely rare so far, but suggest a range of density
  profiles across Seyferts. We resolve a weeks-long absorption event
  in the Seyfert NGC 3227.}{We examine six \suzaku and twelve \swift\
  observations from a 2008 campaign spanning 5 weeks. We use a
  model accounting for the complex spectral interplay of
  three differently-ionized absorbers. We perform time-resolved
  spectroscopy to discern the absorption variability behavior. We also
  examine the IR-to-X-ray spectral energy distribution (SED) to test
  for reddening by dust.}{The 2008 absorption event is due to
  moderately-ionized ($\log \xi\sim 1.2$--$1.4$) gas covering 90\% of
  the line of sight. We resolve the density profile to be highly
  irregular, in contrast to a previous symmetric and centrally-peaked
  event mapped with RXTE in the same object. The UV data do not show
  significant reddening, suggesting that the cloud is dust-free.}{The
  2008 campaign has revealed a transit by a filamentary,
  moderately-ionized cloud of variable density that is likely located
  in the BLR, and possibly part of a disk wind.}


\maketitle

\section{Introduction}\label{sec:intro}
X-ray spectroscopy allows us to probe circumnuclear matter obscuring
the direct line of sight to the supermassive black holes (SMBHs) that
power AGN. This material can be part of a dusty torus surrounding the
nucleus \citep{Antonucci1993}, of the Broad Line Region (BLR), or a
disk outflow. Variable absorption has been found for a number of AGN,
e.g., a sample of 12 Seyfert 1.5 galaxies with inclination angles
comparable to the torus opening angle, which included
\object{NGC\,3227} \citep{Beuchert2013} as well as other AGN with
intermediate optical classifications such as \object{NGC\,4051}
\citep{Guainazzi1998}, \object{MCG$-$6-30-15} \citep{McKernan1998}, or
\object{NGC\,3516} \citep{Turner2008}.  \citet{Risaliti2002} study an
X-ray selected set of 25 Seyfert~2 galaxies and find soft X-ray
variability on both long and short time-scales. Short-term
($\sim$1\,d) absorption events have been detected, e.g., for
\object{Mrk~766} \citep{Risaliti2011}, \object{NGC~5506}
\citep{Markowitz2014}, \object{NGC~4388} \citep{Elvis2004} and
\object{NGC\,1365}
\citep{Risaliti2007,Risaliti2009b,Risaliti2009a}. Long-term events
($\geq 7\,\mathrm{d}$) have been found for, e.g., \object{Cen\,A}
\citep{Rivers2011a}, \object{Fairall~9} \citep{Lohfink2012}, or
\object{NGC\,3227} in 2000/2001 \citep{Lamer2003}, thanks to the Rossi
X-ray Timing Explorer (\rxte) monitoring.

These absorption events can be explained by transits of discrete
clouds or clumps of gas across the line of sight to the central X-ray
continuum source
\citep[e.g.,][]{Risaliti2002,Lohfink2012,Markowitz2014}. They support
a new generation of ``clumpy torus'' models
\citep{Elitzur2007,Nenkova2002,Nenkova2008a,Nenkova2008b}. In these
models, clouds are typically concentrated towards the equatorial plane
but with a soft-edge angular distribution, usually on near-Keplerian
orbits, and possibly embedded in a tenuous inter-cloud medium
\citep{Stalevski12}. Outside the dust sublimation radius, which is
typically several light-weeks away from the SMBH, the presence of such
a clumpy component is supported by fits to the infrared spectral shape
\citep{AsensioRamos2009,AlonsoHerrero2011}. Closer to the SMBH the
population of dusty clouds may transition to the dust-free clouds that
comprise the BLR \citep{Elitzur2007}. A clumpy X-ray absorbing medium
located at distances commensurate with the BLR -- and possibly
identified as BLR clouds themselves -- has also been suggested
\citep{Risaliti2009b,Risaliti2011}.
\citeauthor{Arav1998}'s~\citeyear{Arav1998} investigation of optical
line profiles in NGC~4151 also supports such a distribution of
discrete BLR clouds, with a large number of rather small clouds. From
an observational point of view, however, the mid-IR emission only
probes matter outside the dust sublimation zone, while X-ray
absorption probes the full radial range. Many of the short-term X-ray
absorption events have been interpreted as BLR clouds
\citep{Risaliti2007}, while the longer term events found, e.g., by
\citet[][durations of more than a few days to more than a
  year]{Markowitz2014} were inferred to be due to clouds residing in
the outer BLR or the inner dusty torus.

When adequate data have been available, i.e., sustained sampling on
timescales longer than the eclipse duration that resolves the
eclipses, one has been able to pinpoint their ingress and egress, and
also to obtain time-resolved information on the column density profile
$N_\mathrm{H}(t)$ along the transverse direction. Only a few density
profiles have been resolved: The events in NGC~3227 and Cen~A
mentioned above featured symmetric, non-uniform, and centrally-peaked
column density profiles. An event in Mkn~348 may also fall into this
category \citep{Akylas2002}. \citet{Maiolino2010} report comet-shaped
clouds, with dense ``heads'' and less-dense ``tails'', in NGC 1365.
\citet{Markowitz2014} report a doubly-peaked absorption event in
NGC~3783. These results indicate a broad variety of profile shapes,
hinting at a range in cloud origins and/or the physical mechanisms
that shape clouds.

To add complexity to the interpretation of X-ray spectra, a
significant number of Seyferts also show evidence for a complex
interplay of differently ionized layers of warm absorbing gas along
the line of sight \citep[e.g.,][]{Blustin2005,Turner2008}. This gas
likely originates in the accretion disk, the BLR, or the inner torus
\citep{George1998,Krolik2001,Reynolds1995}. Blue-shifted absorption
features indicate that some warm absorbers are outflowing along the
line of sight \citep{Blustin2005} and may be launched from the inner
accretion disk \citep{Krongold2007}. The observed warm absorbers could
be part of stratified outflows \citep{Tombesi2013}. Such disk winds
can be launched by magneto-hydrodynamical (MHD) forces
\citep{BP1982,Contopoulos1994,Fukumura2010}.

The Seyfert 1.5 AGN NGC~3227 was subjected to sustained monitoring
with \rxte from 1999-Jan-02 to 2005-Dec-04 \citep{Uttley2005}. The
monitoring revealed two eclipse events: an $\sim$80-day event in
2000/2001 \citep{Lamer2003} and a 2--7\,day event in 2002
\citep{Markowitz2014}. The estimated distances from the central engine
are tens of light-days (ld). Given the estimated location of the BLR
from the SMBH, at 2--19\,ld \citep{Peterson2004,Landt2008}, and the
inner edge of the IR-emitting torus as determined by reverberation
mapping, $\sim20\,\mathrm{ld}$ \citep{Suganuma2006}, these clouds are
likely in the inner dusty torus, or at least the outermost BLR.

In this paper we present an additional, intriguing, absorption event
in NGC~3227 found in quasi-simultaneous \suzaku and \swift\ data from
2008. The column density profile turns out to be highly irregular. In
Sect.~\ref{sec:obsdatared} we present the data reduction while
Sect.~\ref{sec:spectra} revisits the two already published \xmm\
observations in order to define a model that we use to simultaneously
fit the archival \suzaku observations.  The X-ray and UV data
provided by \swift\ are also analyzed. In Sect.~\ref{sec:origin} we
discuss the properties and location of the absorbing cloud and its
possible origin in light of these observations.

\section{Observations and Data Reduction}\label{sec:obsdatared}
\subsection{Observations}
\label{subsec:observations}
\begin{figure}
  \resizebox{\hsize}{!}{\includegraphics{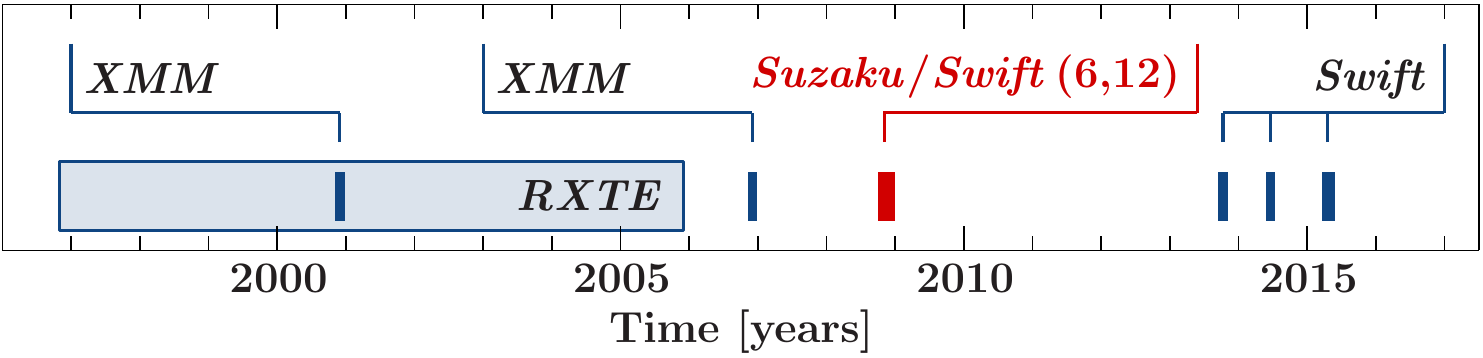}}
  \caption{Time-line including all observations of NGC~3227 discussed
    here. The blue region shows the time-range of sustained \rxte\
    monitoring from 1999-Jan-02 to 2005-Dec-04 \citep{Uttley2005}. The
    observations by \suzaku and \swift\ from the 2008 campaign are
    displayed in red, the times of the \xmm and recent \swift\
    observations are shown in blue.}
  \label{fig:timeline}
\end{figure}

Figure~\ref{fig:timeline} provides an overview over all archival
observations of NGC~3227 that we consider. NGC~3227 was the subject of
a 35\,d long \suzaku and \swift\ monitoring campaign between 2008
October 2 and 2008 December 03. The campaign consists of six
\suzaku and twelve nearly simultaneous \swift\ observations
(Table~\ref{tab:suzswift_obs}). Outside of this campaign, further
archival data with good soft X-ray coverage are provided by two
\xmm observations, one from 2000 November 28--29 at the onset of the
2000/2001 absorption event \citep[obsid: 0101040301, exposure time
  post screening 32\,ks;][]{Lamer2003} and a 97\,ks observation on
2006 December 3--4 \citep[obsid: 0400270101;][]{Markowitz2009} during
a state with intrinsically high flux and relatively low absorption.
The observation from 2006 yielded reflection grating spectrometer
(RGS) data of sufficient signal-to-noise for a detailed analysis of the
absorbers. Recent \swift\ pointings from 2013 October 20 until 2015
May 02 are also included in our analysis.

For the reduction of these data we used \textsc{HEASOFT} v.\ 6.15.1
and \xmm \textsc{SAS} v.\ 13.5.0. The spectral analysis is performed
with the Interactive Spectral Interpretation System
\citep[\textsc{ISIS};][]{Houck2000}. Uncertainties on spectral fit
parameters correspond to the 90\% confidence level for one interesting
parameter. Throughout the paper, we use the systemic redshift of $z =
0.00386$ \citep{Vaucouleurs1991} and a Galactic equivalent hydrogen
column of $N_\mathrm{H}=1.99\times 10^{20}\,\mathrm{cm}^{-2}$
\citep{Kaberla2005}. The black hole mass is assumed to be $1.75\times
10^{7}\,M_{\astrosun}$, the average of various measurements collected
at the AGN Black Hole Mass
Database\footnote{\url{http://www.astro.gsu.edu/AGNmass/}}.
Luminosities are calculated using a luminosity distance of 20.3\,Mpc
\citep{Mould2000}. We perform the $K$-correction according to
\citet{Ghisellini2009}. Neutral absorption is modeled with
\textsc{tbnew}, an improved version of the absorption model of
\citet{Wilms2000}, using cross sections from \citet{Verner1996} and
abundances from \citet{Wilms2000}. Ionized absorption is modeled using
the \textsc{zxipcf} model, which is based on XSTAR tables
\citep{Kallman2001}. In this model, following \citet{Tarter1969} the
ionization parameter, $\xi$, for a medium at distance $R$ from a
source of luminosity $L_\mathrm{ion}$ is defined as
\begin{equation}\label{eq:def_ion}
\xi=L_\mathrm{ion}/(n_\mathrm{H}\,R^{2})
\end{equation}
where $n_\mathrm{H}$ is the hydrogen number density of the absorber.
\begin{table*}
  \caption{\suzaku and \swift\ observations in 2008 with their
    screened exposure times. Counts given are in the energy bands
    given in Sect.~\ref{sec:obsdatared}. The check-symbol ($\surd$)
    denotes observations that are used in the data analysis, while
    observations labeled with a cross (\ding{55}) are excluded (see
    Sect.~\ref{subsec:datared_swift} for details).}
\label{tab:suzswift_obs}
  \centering
  \scriptsize
  \begin{tabular}[ht]{llllllllllllll}
    \toprule
    \multicolumn{7}{l}{\suzaku XIS} & \multicolumn{7}{l}{\swift\ XRT}\\
    \midrule
    abbrv. &obsid     & det   & time & exp [ks] & cnts [$\times 10^4$] &  & abbrv. & obsid & mode & date & exp [ks] & cnts &\\
    \midrule
    Suz~1 & 703022010 & XIS 0 & 2008-10-28 & 58.9 & 7.8   & $\surd$ & Sw~1a & 00037586001 & pc & 2008-10-28 & 1.0 & 543 & $\surd$\\
             & & XIS 1 &                  & & 8.6 & $\surd$ & Sw~1b & 00037586002 & pc & 2008-10-29 & 2.0 & 2386 & \ding{55}\\
             & & XIS 3 &                   &       & 8.0 & $\surd$ &         &    &    &            &       &      & \\
             & & HXD   &                   & 47.9       & 3.4 & $\surd$ &         &    &    &            &       &      & \\
    Suz~2 & 703022020 & XIS 0 & 2008-11-04 & 53.7 & 2.2    & $\surd$ & Sw~2a & 00031280001 & wt & 2008-11-04 & 1.0 & 255 & $\surd$\\
            &  & XIS 1 &            & & 2.2 & $\surd$ & Sw~2b &00031280002 & wt & 2008-11-05 & 2.2 & 497 & $\surd$\\
            &  & XIS 3 &            & & 2.2 & $\surd$ &        &     &    &            &       &      & \\
            &  & HXD   &            & 46.4 & 3.1 & $\surd$ &         &    &    &            &       &      & \\
    Suz~3 & 703022030 & XIS 0 & 2008-11-12 & 56.6 & 3.3    & $\surd$ & Sw~3a & 00031280003 & wt & 2008-11-12 & 1.0 & 291 & $\surd$\\
            &  & XIS 1 &            & & 3.4 & $\surd$ & Sw~3b &00031280004 & pc & 2008-11-13 & 2.2 & 702 & $\surd$\\        
            &  & XIS 3 &            & & 3.4 & $\surd$ &        &     &    &            &       &      & \\
            &  & HXD   &           & 46.6  & 3.2 & $\surd$ &        &     &    &            &       &      & \\
    Suz~4 & 703022040 & XIS 0 & 2008-11-20 & 64.6 & 1.5    & $\surd$ & Sw~4a & 00031280005 & pc & 2008-11-21 & 2.1 & 280 & $\surd$\\        
            &  & XIS 1 &           &  & 1.8 & $\surd$ & Sw~4b & 00031280006 & pc & 2008-11-22 & 1.9 & 256 & $\surd$\\        
            &  & XIS 3 &           &  & 1.5 & $\surd$ &       &      &    &            &       &      & \\
            &  & HXD   &           & 43.4  & 2.9 & $\surd$ &       &      &    &            &       &      & \\
    Suz~5 & 703022050 & XIS 0 & 2008-11-27 & 79.4 & 3.7    & $\surd$ & Sw~5a & 00031280007 & pc & 2008-11-25 & 2.0 & 429 & $\surd$\\        
            &  & XIS 1 &             & & 3.8 & $\surd$ & Sw~5b &00031280008 & pc & 2008-11-27 & 1.9 & 252 & $\surd$\\        
            &  & XIS 3 &             & & 3.7 & $\surd$ &       &      &    &            &       &      & \\
            &  & HXD   &             & 37.2 & 2.6 & $\surd$ &       &      &    &            &       &      & \\
    Suz~6 & 703022060 & XIS 0 & 2008-12-02 & 51.4 & 1.7    & $\surd$ & Sw~6a &00031280009 & pc & 2008-12-02 & 0.3 & 50 & \ding{55}\\        
            &  & XIS 1 &            &  & 1.8 & $\surd$ & Sw~6b &00031280010 & pc & 2008-12-03 & 1.7 & 264 & $\surd$\\        
            &  & XIS 3 &            &  & 1.7 & $\surd$ &       &      &    &            &       &      & \\
            &  & HXD   &            & 36.5 & 2.3 & $\surd$ &       &      &    &            &       &      & \\
  \end{tabular}
\end{table*}

\subsection{\xmm}\label{subsec:datared_xmm}
Both \xmm observations were taken in full-frame mode of the EPIC-pn
camera \citep{Strueder2001, Turner2001}. After creating calibrated
event lists with filtered hot and bad pixels, events in 10--12\,keV
are screened for enhanced rates due to particle flaring. As no pile-up
is evident, we extract 0.3--10\,keV spectra from all counts within
$40''$ of the central source position of NGC~3227. The same angular
radius of the extraction region is used to extract the background from
a position $\sim$6\arcmin off-source on the same CCD-chip.

\subsection{\suzaku}\label{subsec:datared_suzaku}
We extract both the data of the X-ray Imaging Spectrometer
\citep[XIS;][]{Koyama2007} and the Hard X-ray Detector
\citep[HXD;][]{Takahashi2007}. We use data taken by the front
(XIS~0,3) and back-illuminated (XIS1) chips in the $3\times3$ and
$5\times5$ editing modes.  We reprocessed the unfiltered event lists
by applying the newest calibration available and screened the data
with default parameters.  We then perform an attitude correction with
\textsc{aeattcor2} based on the $3\times 3$ mode which comprises the
bulk of data. The resulting spectra of the $3\times 3$ and $5\times 5$
modes are merged using \textsc{mathpha} for each XIS. Spectra were
extracted from circular regions of $\sim$93\arcsec\ radius and
centered on the point source. We normalize all fluxes with respect to
XIS0. Our fits find that the flux normalization of the XIS1 and XIS3
spectra deviates by about 5\% from that of the XIS0 spectrum,
consistent with the \suzaku ABC guide
(\url{http://heasarc.gsfc.nasa.gov/docs/suzaku/analysis/abc/}),
version~5.0. We perform simultaneous fits of observations Suz~2--Suz~6
using the SimFit routines of \texttt{ISIS} \citep{kuehnel2015a}, and
rebin their spectra to a combined minimum signal-to-noise ratio (S/N)
of 25. Observation Suz~1 is fitted separately and binned to a minimum
S/N of 18. In both cases the binning does not exceed the energy
resolution of the XIS detectors of 150\,eV at $\sim$6\,keV in
2008\footnote{\url{https://heasarc.gsfc.nasa.gov/docs/astroe/prop_tools/suzaku_td/node10.html}}. Due
to calibration uncertainties caused by the Si- and Au-edges we exclude
data in the 1.72--1.88\,keV and 2.19--2.37\,keV energy bands. For
Suz~1 the first interval is extended to 1.5--1.88\,keV because of
further insufficient calibration just above 1.5\,keV.

The non-imaging HXD-PIN data are extracted for the whole field of view
of $34\arcmin \times 34\arcmin$. For a list of total counts and good
exposure times see Table~\ref{tab:suzswift_obs}. The flux
normalization of the HXD with respect to XIS0 has a large uncertainty
because of the low signal to noise ratio of the HXD data, but is
consistent with what is expected for the current calibration. We
therefore fix the HXD flux normalization constant to its nominal value
of 1.16.

\subsection{\swift}\label{subsec:datared_swift}
The X-ray Telescope (XRT) on board \swift\ \citep{Burrows2005} provides
imaging and spectroscopic capabilities for the energy range
0.5--10\,keV. Except for observations Sw~2a,b and Sw~3a, which were
taken in windowed-timing (WT) mode, all observations were performed in
photon-counting (PC) mode. For the PC mode we allow grades from 0
to~12. Only grade 0 events are extracted for the WT mode. We extract
source counts within $30'$ of the source position. The background is
extracted from circular regions of at least $60''$ radius on a
position on the chip which is free of background sources. We exclude
observation Sw~1b from our analysis because of an excess above 7\,keV
of unknown origin. Observation Sw~6a was also ignored due to an
insufficient number of detected counts.

The Ultraviolet and Optical Telescope (UVOT) on board \textsl{Swift}
observes with up to six filters \citep[UVW2, UVM2, UVW1, U, B, and
  V]{Roming2005}. The UVOT image data have been summed using
\textsc{uvotimsum} (V.~24Jan2014\_V6.15.1). A circular region of $5''$
around the source position was used for the extraction of counts of
the central core component. The background was defined as an annulus
around the source with $13''$ and $26''$ inner and outer radius, to
minimize contributions from the host galaxy. The resulting files were
converted into pha files using the \textsc{uvot2pha} task and
de-reddened as discussed in Sect.~\ref{sec:analysis_swift}.

\section{X-Ray Spectral Analysis}\label{sec:spectra}
The main focus of the data analysis lies on the simultaneous \suzaku\
and \swift\ data describing an additional absorption event.  In the
following, we first build up a spectral baseline model using the
\suzaku data. With help of this model, we re-visit the two archival
\textit{XMM-Newton} observations and finally model each set of
\suzaku and \swift\ observations simultaneously.

\subsection{A baseline model based on \suzaku data}
Before we start to develop a baseline spectral model for NGC~3227, we
briefly review the previous attempts to describe the spectrum.
\citet{Lamer2003} describe the 2000 \xmm observation with a power law
absorbed by neutral or lowly-ionized material (hereafter \WA{1}) and
an unabsorbed power law. The slope of the absorbed power law was
difficult to constrain. As shown by \citet{Lamer2003}, this model is
highly degenerate in $\xi$ (two orders of magnitude) and
$N_\mathrm{H}$. We find that a lower degree of degeneracy can be
obtained by assuming a partial covering scenario which yields
comparable statistics. The lack of sufficient RGS data and the
relatively low signal-to-noise did not allow \citet{Lamer2003} to
constrain more absorbers.

In contrast, the 2006 \xmm observation caught the source in a typical
hard X-ray flux and spectral state, i.e., unabsorbed by moderately
Compton-thick gas. The high soft X-ray flux and the long exposure time
allowed \citet{Markowitz2009} to obtain a high S/N RGS spectrum. They
constrained two layers of ionized absorption, one intermediately
ionized, $\log \xi \sim 1.45$ and one highly ionized, $\log \xi \sim
2.93$ absorber. We call these \WA{2} and \WA{3} in the following.
\citet{Markowitz2009} also detected an absorber with a small neutral
column, which is similar to \WA{1} seen in 2000 by \citet{Lamer2003}.

In deriving the baseline model for our analysis, we use the high
signal to noise \suzaku data and assume from now on that \WA{1},
\WA{2}, and \WA{3} -- assumed to represent physically distinct
absorbing media with different levels of ionization -- are present in
all \xmm, \suzaku, and \swift observations.  The assumption that WA1,
WA2 and WA3 are present in all observations and stable over durations
of many years is a simplifying one; in some cases, a given absorber
may not be statistically required in the fits, particularly if the
column densities of the other absorbers are such that they dominate
the spectral shape.  In addition, as demonstrated below, we observe a
wide range in column densities in WA1 and WA2 across spectra taken
several years apart.  A simple assumption is that each absorber
remains present in all observations, and has varied primarily in
column density, and not ionization state. However, we cannot exclude
the possibility of some observations capturing an additional
physically distinct absorber with similar ionization level moving into
the line of sight.

\begin{figure}
  \resizebox{\hsize}{!}{\includegraphics{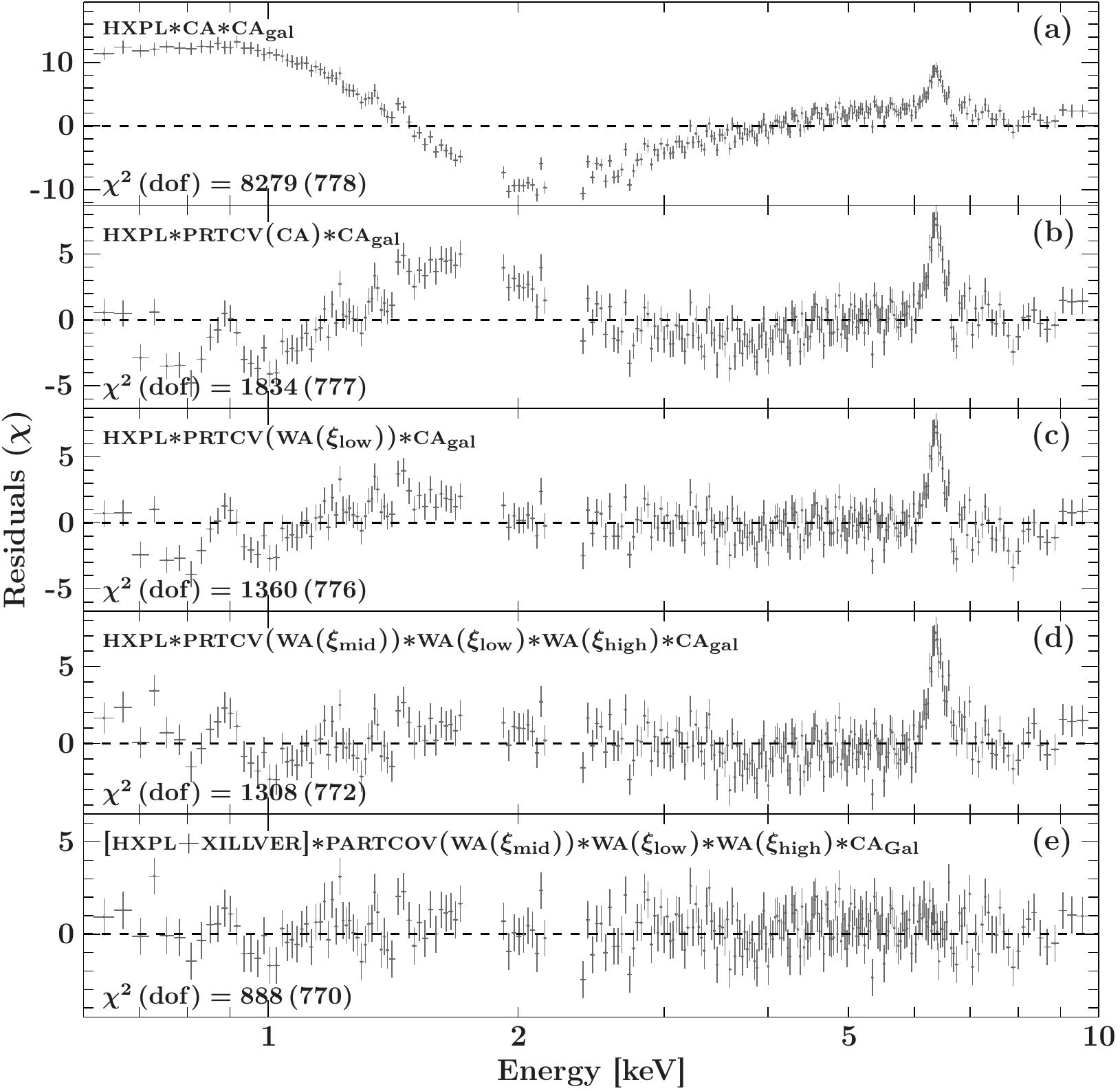}}
  \caption{Example of the steps followed to find the best-fit baseline
    model on the \suzaku spectra (we use observation Suz~3 in this
    example). \textbf{a} Fit to the spectrum with a
    single power law absorbed by neutral gas. \textbf{b} The neutral
    absorber is partially covering the incident power law. \textbf{c}
    Due to characteristic residuals we replace the solely neutral
    absorber by an ionized one and \textbf{d} and two more warm
    absorbers leading to a stratified composite of differently ionized
    absorbers. \textbf{e} The remaining residuals in the soft band and
    around 6.4\,keV are flattened when finally adding a reflected
    power law in panel.}
  \label{fig:703022030_bottom_up}
\end{figure}

Figure~\ref{fig:703022030_bottom_up} shows the steps followed to find
the best-fit baseline model on the 2008 \suzaku spectra motivated by
\citet{Lamer2003} and \citet{Markowitz2009}. We use observation Suz~3
in this example, which has the best combination of count rate and
strong absorbing column density. The $\chi^2$ of each fit is labeled
in the figure and shows the gradual improvement of the model. In
Fig.~\ref{fig:703022030_bottom_up}a the high-energy X-ray power law
(HXPL) absorbed by one layer of neutral absorption (CA). Apart from
the prominent Fe~K$\alpha$ line around 6.4\,keV, the strong residuals
below 4\,keV suggest a partial covering scenario similar to that seen
in the 2000 \xmm data. Assuming partial absorption by a neutral
medium results in the residuals shown in
Fig.~\ref{fig:703022030_bottom_up}b. Below 3\,keV the residuals still
show strong positive and negative excesses. Replacing the neutral
absorber by a warm absorbing layer of very low ionization results in
substantially flattened residuals
(Fig.~\ref{fig:703022030_bottom_up}c). We identify this component with
\WA{1}.

While this model is able to describe the smooth turnover of the
partial coverer, some residuals remain. Based on the 2006 \xmm\
observation, we model these residuals by including two additional
layers of ionized absorption, \WA{2} and \WA{3}
(Fig.~\ref{fig:703022030_bottom_up}d). Here, following the RGS
analysis of \citet{Markowitz2009}, the absolute redshift of \WA{3} is
fixed at $-0.00302$. The fit requires the intermediately ionized
absorber \WA{2} ($\logxi\sim 0.5$) to be dominant and partially
covering.

Leftover residuals in the final continuum fit are line-like features
in the soft-band around 0.9\,keV, which are most likely due to
\ion{Ne}{ix} emission, and the strong signature of a
Fe~K$\alpha$/Fe~K$\beta$ complex between 6.4 and 7.1\,keV. These soft
and hard components can either be modeled individually using empirical
emission lines, or by adding ionized reflection \citep[modeled with
  \textsc{xillver};][]{Garcia2013}, where we require that the reflected
continuum has the same shape as the primary continuum. Both approaches
describe all of the excess components equally well and without
appreciably changing the continuum parameters. Because of the unique
physical interpretation and the smaller number of free parameters, we
continue the modeling by using \textsc{xillver}. The resulting model
describes the data very well (Fig.~\ref{fig:703022030_bottom_up}e). 

\subsection{Re-Visiting the \xmm spectra with the baseline model}
\begin{figure}
  \resizebox{\hsize}{!}{\includegraphics{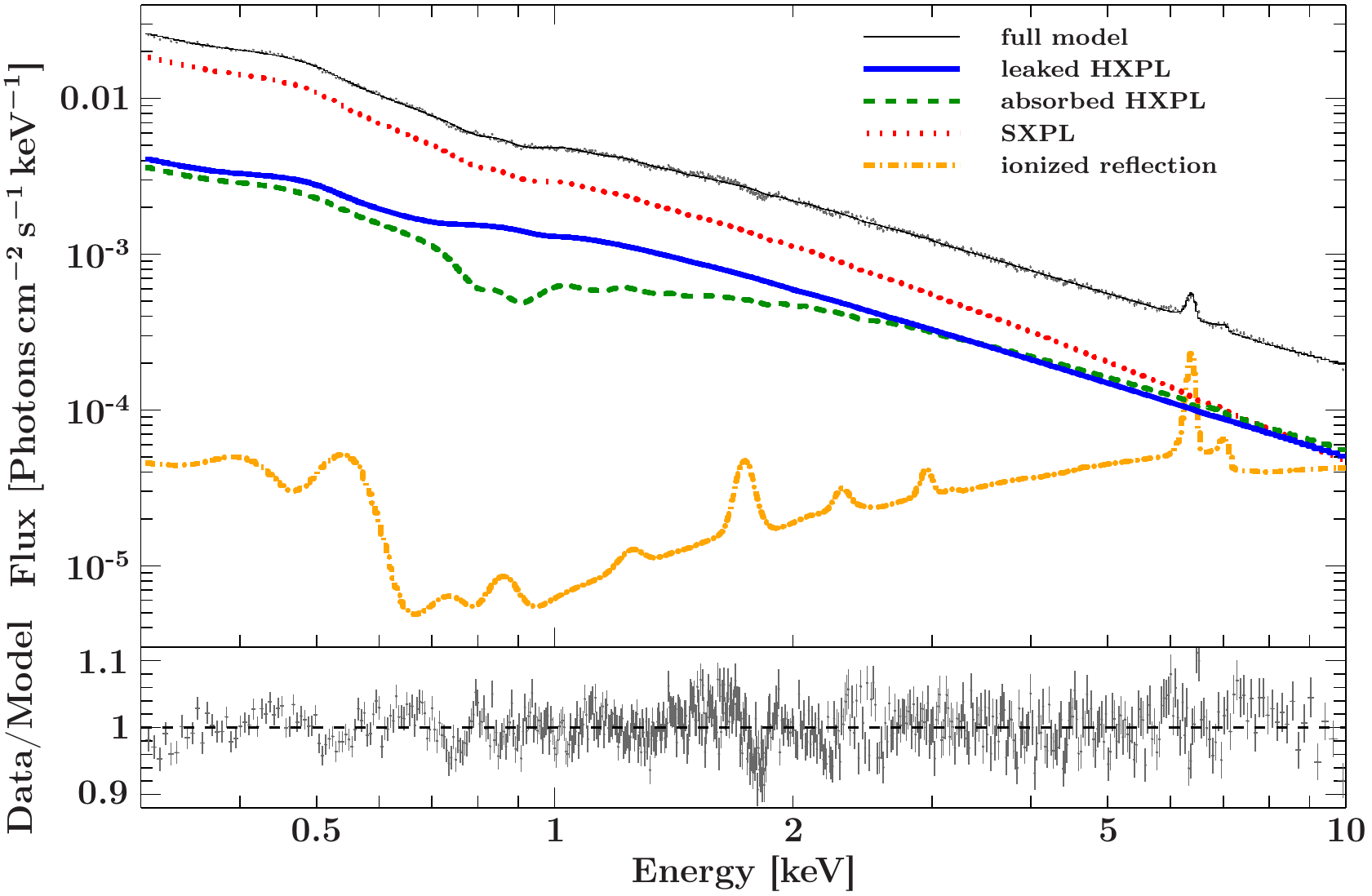}}\hfill
  \resizebox{\hsize}{!}{\includegraphics{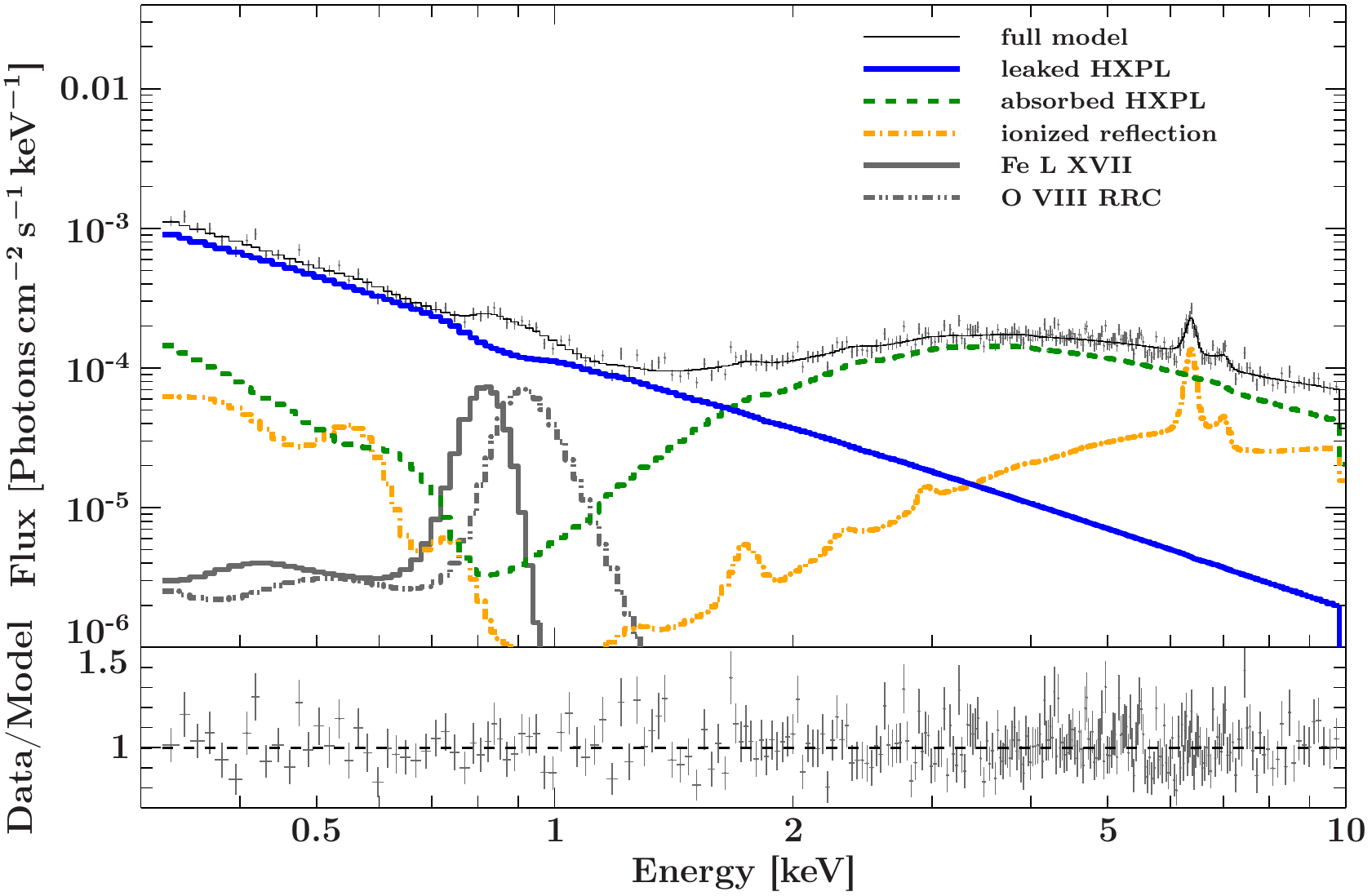}}
  \caption{The 2006 (top) and 2000 (bottom) \xmm spectra, the best
    fit model components, and the complete model shown with thick and
    thin solid lines, respectively. Both spectra are described by the
    baseline model that comprises a high energy X-ray power law with a
    reflection component and absorption by three layers of lowly,
    intermediately and highly ionized gas as well as a Galactic
    column. The source is partially covered by \WA{2} for the 2006 and
    by \WA{1} for the 2000 observation. The data from 2006 also
    require a steep soft excess. The sharp residuals at
    $\sim$1.85\,keV are likely due to calibration uncertainties around
    the Si~K edge.}
  \label{fig:xmm_models}
\end{figure}

We now test whether the baseline model also describes the \xmm\
observations that have been used to motivate this model and its
absorbers. The spectra and model components of the 2000 and 2006 \xmm\
observations are shown in Fig.~\ref{fig:xmm_models}. The continuum is
described by an absorbed high-energy power law, leaked emission, and
unblurred ionized reflection
components. Table~\ref{tab:model_0400270101} and
Table~\ref{tab:model_0101040301} list the best-fit parameters.

Because of the inability to constrain the covering fraction of the
mildly (\WA{1}) and highly (\WA{3}) ionized absorbers, we
always require them to fully cover the compact source, while their
column densities and ionization parameters are all left free to vary.

\begin{figure}
  \resizebox{\hsize}{!}{\includegraphics{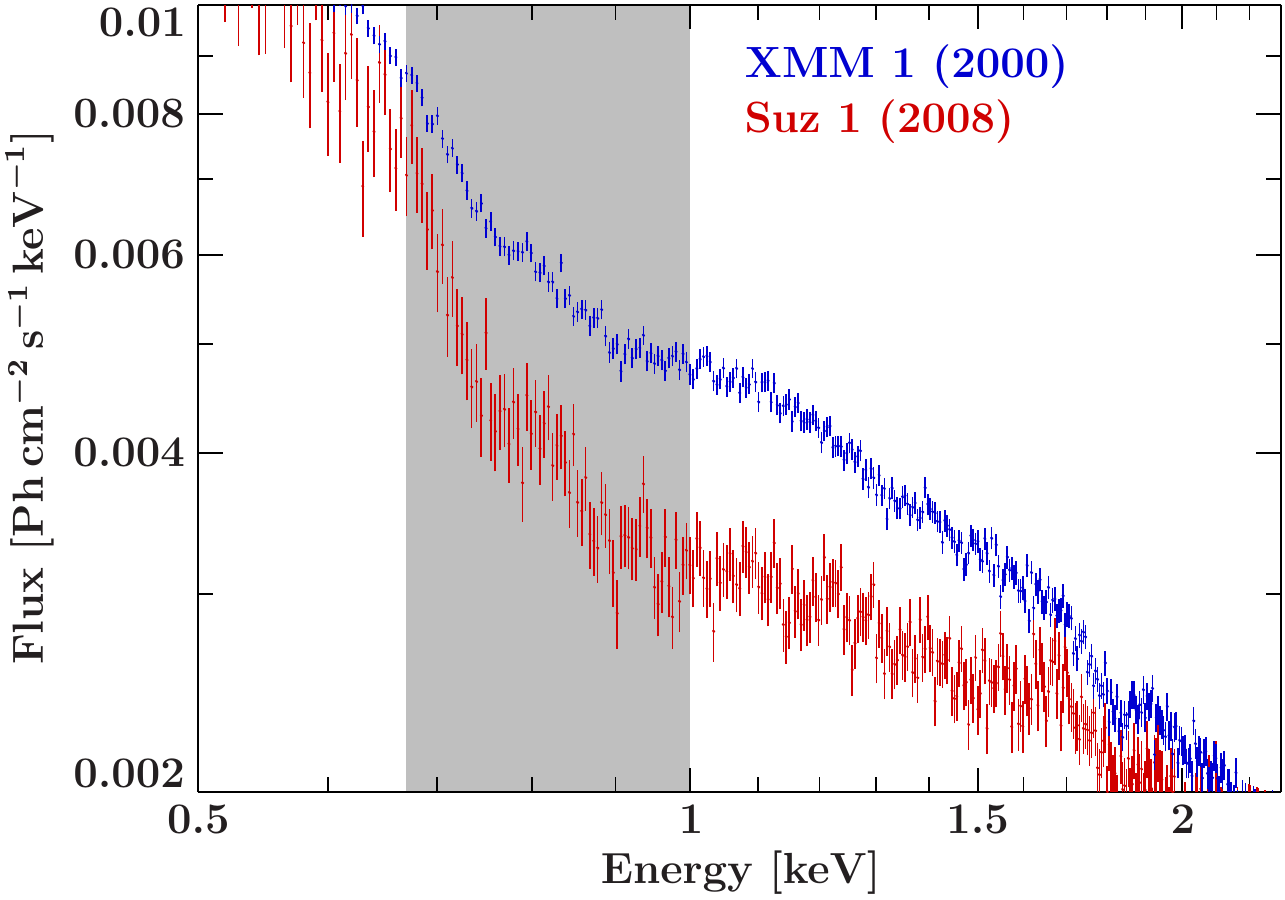}}
  \caption{Comparison of the 2006 \xmm and Suz~1 observations in the
    0.5--2.3\,keV range. The gray area marks the energy range
    comprising two modest absorption features that are due to an
    absorber of similar $\xi$, i.e., \WA{2}.}
  \label{fig:xmm_suz}
\end{figure}

The continuum as well as the Fe~K~$\alpha$/$\beta$ lines of the 2006
\xmm observation are well fitted with a combination of an incident
power law and ionized reflection component
\citep[\texttt{xillver}]{Garcia2013}. The two modest absorption
dips at around 0.75\,keV and 0.9\,keV require \WA{2} to have $\log\xi
\sim 2$ and to cover the X-ray source by $\sim$50\%.
Figure~\ref{fig:xmm_suz} shows the 0.5--2.3\,keV unfolded spectra of
both the Suz~1 observation from 2008 and the \xmm observation from
2006. The two absorption features enclosed by the gray shaded region
are found at about the same energy in both spectra, arguing for \WA{2}
as a common origin. The similar spectral shape prefers \WA{2} to be
partial covering in both observations. In contrast to the 2000 data, a
steep soft excess is seen in 2006, which we describe by a soft power
law (SXPL), consistent with the results of \citet{Markowitz2009}. The
fit converges to a statistic of $\chi^{2}/\mathrm{dof}=1851/1542$.

While the data from 2006 are described well with the partial covering
absorber \WA{2}, the 2000 data require a dominant partial covering
column of lowly-ionized gas (\WA{1}). Given the strong X-ray
absorption, it is not clear if a soft excess is present in the
spectrum. If our partial-covering model is correct and the continuum
below $\sim$2\,keV is dominated by ``leaked'' emission, then the soft
excess ($\Gamma \sim 3$) must have had a negligible presence. Given
the low signal-to-noise ratio of the data, with the exception of the
column of \WA{2} the parameters of the two highest ionized absorbers
\WA{2} and \WA{3} cannot be constrained, and are assumed to be equal
to those found in the 2006 data. The emission feature at
$\sim$0.88\,keV is described by a narrow unresolved Gaussian for
\ion{Fe~L}{xvii} at 0.826\,keV and an \ion{O}{viii} radiative
recombination continuum (RCC) at 0.871\,keV. We find the best-fit
statistic to be $\chi^{2}/\mathrm{dof}=901/847$, which is consistent
with the one found by \citet{Lamer2003}.

\begin{table}
  \caption{Continuum parameters of the spectral fit of the baseline
    model to the relatively unabsorbed 2006 \xmm observation. Fixed
    parameters are marked with an asterisk. We assume that the
    incident hard X-ray power law (HXPL) of the ionized
    reflection and the continuum power law are identical. 
    To reduce model degeneracies, we freeze the photon index 
    to the value found by \citet{Markowitz2009}, who find \WA{3} to be outflowing. 
    We therefore also freeze its absolute redshift to the according value.
    The soft excess is modeled by a
    power law (SXPL).}
\label{tab:model_0400270101}
  \centering\small
  \renewcommand{\arraystretch}{1.3}
\begin{tabular}{lll}
  \hline\hline
\multicolumn{3}{l}{\textsc{[pow+xillver]$\ast$}}\\
\multicolumn{3}{l}{\textsc{zxipcf(wa$_1$)$\ast$zxipcf(wa$_2$)$\ast$zxipcf(wa$_3$)$\ast$tbnew\_simple(local)}}\\
\hline
  mod. comp. & parameter & value\,$\pm$\,uncert.\\
  \hline
\multicolumn{3}{c}{0400270101: $\chi^2$\,(dof) = 1341.1\,(953)}\\
\midrule
SXPL & norm & $\left(4.1^{+0.5}_{-0.6}\right)\times10^{-3}$\\
 & $\Gamma$ & 1.57$^\ast$\\
HXPL & norm & $\left(6.2^{+0.7}_{-0.5}\right)\times10^{-3}$\\
 & $\Gamma$ & $2.099^{+0.041}_{-0.021}$\\
Ion. Refl. & norm & $\left(2.670^{+0.004}_{-0.757}\right)\times10^{-4}$\\
 & $\log\xi$\,($\mathrm{erg}\,\mathrm{cm}\,\mathrm{s}^{-1}$) & $0.17^{+0.16}_{-0.17}$\\
 & $Z_\mathrm{Fe}$ & $0.50^{+0.04}_{-0.00}$\\
 & $z$ (absolute) & 0.00386$^\ast$\\
WA 1 & $N_\text{H}$\,($10^{22}\mathrm{cm}^{-2}$) & $0.1391^{+0.0006}_{-0.0013}$\\
 & $\log\xi$\,($\mathrm{erg}\,\mathrm{cm}\,\mathrm{s}^{-1}$) & $-0.521^{+0.018}_{-0.017}$\\
 & f$_\text{cvr}$ & 1$^\ast$\\
 & $z$ (absolute) & 0.00386$^\ast$\\
WA 2 & $N_\text{H}$\,($10^{22}\mathrm{cm}^{-2}$) & $2.75\pm0.29$\\
 & $\log\xi$\,($\mathrm{erg}\,\mathrm{cm}\,\mathrm{s}^{-1}$) & $2.01^{+0.06}_{-0.10}$\\
 & f$_\text{cvr}$ & $0.533^{+0.023}_{-0.045}$\\
 & $z$ (absolute) & 0.00386$^\ast$\\
WA 3 & $N_\text{H}$\,($10^{22}\mathrm{cm}^{-2}$) & $0.11^{+0.22}_{-0.07}$\\
 & $\log\xi$\,($\mathrm{erg}\,\mathrm{cm}\,\mathrm{s}^{-1}$) & $3.10^{+0.23}_{-0.30}$\\
 & f$_\text{cvr}$ & 1$^\ast$ \\
 & $z$ (absolute) & -0.00302$^\ast$\\
CA & $N_\text{H,Gal}$\,($10^{22}\mathrm{cm}^{-2}$) & 0.0199\\
\bottomrule
\end{tabular}
\end{table}

\begin{table}
  \caption{Continuum parameters of the spectral fit of the derived
    baseline model to the absorbed 2000 \xmm observation. Fixed
    parameters are marked with an asterisk, parameters adopted from
    the fit to the 2006 \xmm observation with $\dagger$. The photon
    index of the reflection component is tied to the one of the high
    energy power law. The model also includes an unresolved Gaussian
    line at $E_\mathrm{c}=0.826\,\mathrm{keV}$ identified as
    \ion{Fe~L}{xvii} and a \ion{O}{viii} radiative recombination continuum (RRC)
    with an edge energy of 0.871\,keV.}
  \label{tab:model_0101040301}
\centering\small
\renewcommand{\arraystretch}{1.3}

\begin{tabular}{lll}
  \hline\hline
\multicolumn{3}{l}{\textsc{[pow+xillver]$\ast$}}\\
\multicolumn{3}{l}{\textsc{zxipcf(wa$_1$)$\ast$zxipcf(wa$_2$)$\ast$zxipcf(wa$_3$)$\ast$tbnew\_simple(local)}}\\
  \hline
  mod. comp. & parameter & value\,$\pm$\,uncert.\\
  \hline
  \multicolumn{3}{c}{0101040301: $\chi^2$\,(dof) = 305\,(261)}\\
  \hline
  HXPL & norm & $\left(3.6^{+1.6}_{-0.8}\right)\times10^{-3}$\\
  & $\Gamma$ & $1.89\pm0.20$\\
  Ion.\ Refl. & norm & $\left(2.1^{+1.3}_{-1.8}\right)\times10^{-4}$\\
   & $\log\xi$\,($\mathrm{erg}\,\mathrm{cm}\,\mathrm{s}^{-1}$) & $0.06^{+0.68}_{-0.07}$\\
   & $Z_\mathrm{Fe}$ & $0.60^{+0.37}_{-0.10}$\\
  \WA{1} & $N_\mathrm{H,int}$\,($10^{22}\mathrm{cm}^{-2}$) & $6.6^{+0.6}_{-0.5}$\,\text{$\times\,10^{22}$}\\
  & $\log\xi$\,($\mathrm{erg}\,\mathrm{cm}\,\mathrm{s}^{-1}$) & $0.4^{+0.4}_{-0.5}$\,\text{}\\
  & $f_\mathrm{cvr}$ & $0.959^{+0.012}_{-0.021}$\,\text{}\\
  & $z$ (absolute) & $0.003859^\ast$\\
  \WA{2} & $N_\mathrm{H}$\,($10^{22}\mathrm{cm}^{-2}$) & $0.5^{+1.0}_{-0.5}$\\
  & $\log\xi$\,($\mathrm{erg}\,\mathrm{cm}\,\mathrm{s}^{-1}$) & 2.01$^{\ast,\dagger}$\\
  & $f_\mathrm{cvr}$ & $1^\ast$\\
  & $z$ (absolute) & $0.003859^\ast$\\
  \WA{3} & $N_\mathrm{H}$\,($10^{22}\mathrm{cm}^{-2}$) & 0.2\,\text{$\times\,10^{22}$}$^\ast$\\
  & $\log\xi$\,($\mathrm{erg}\,\mathrm{cm}\,\mathrm{s}^{-1}$) & 3.10$^\ast$\\
  & $f_\mathrm{cvr}$ & $1^\ast$\\
  & $z$ (absolute)& $-0.00302^\ast$\\
Gauss & norm & $\left(3.2^{+1.8}_{-2.0}\right)\times10^{-4}$\\
 & Energy\,(keV) & 0.83$^\ast$\\
 & $\sigma$\,(keV) & 0$^\ast$\\
 & $z$ (absolute) & 0.00386$^\ast$\\
Rec. Edge & norm & $\left(4.7^{+2.8}_{-2.3}\right)\times10^{-4}$\\
 & Energy\,(keV) & 0.87$^\ast$\\
 & $kT$\,(keV) & $0.07^{+0.05}_{-0.04}$\\
  \bottomrule
\end{tabular}
\end{table}
We conclude that the \xmm data of both the absorbed (2000) and
relatively unabsorbed (2006) observations of NGC~3227 can be described
with our baseline model regarding the absorber structure and unblurred
reflection. The spectral variability between both is dominated by the
mildly ionized absorption component \WA{1}.

\subsection{Simultaneous fit to all \suzaku spectra}
The success of the baseline model in describing both the individual
\suzaku data and the \xmm-data suggests that we can use it
to model all 2008 \suzaku spectra simultaneously. The fit results
indicate that the parameters of \WA{1} and \WA{3} stay constant during
the 2008 observational campaign, while the spectral variability is
dominated by WA{2}. We have to model Suz~1 separately, as it does not
share some of the time-independent parameters.
Table~\ref{tab:par_suz} lists the results of the fit.

In more detail, the \textit{time-independent parameters} include the detector
constants, the iron abundance $Z_\mathrm{Fe}$ of the ionized reflector
(a simultaneous fit of Suz~2 to~6 yields $Z_\mathrm{Fe}=2.81\pm 0.17$)
and all parameters of the non-varying ionized absorbers \WA{1} and
\WA{3}. The mildly and highly ionized absorbers \WA{1} and \WA{3} are assumed
to fully cover the central source. \WA{1} has a relatively low column
density of $0.137^{+0.009}_{-0.006}\times10^{22}\,\mathrm{cm}^{-2}$ of
mildly ionized ($\logxi=-0.29^{+0.09}_{-0.13}$) gas for Suz~1 and
$0.068^{+0.025}_{-0.014}\times10^{22}\,\mathrm{cm}^{-2}$
($\logxi=-0.9\pm 0.6$) for the later 5 observations. While both
$\logxi$ are consistent with each other, the column found in Suz~1 is
slightly higher. For \WA{3} the column densities are consistently
around $\sim4\times 10^{22}\,\mathrm{cm}^{-2}$, while the ionization
parameter differs between the single fit of Suz~1 ($\logxi =
3.44^{+0.07}_{-0.05}$) and the other observations, which have
($\logxi=4.17^{+0.16}_{-0.22}$).
\begin{figure}
  \resizebox{\hsize}{!}{\includegraphics{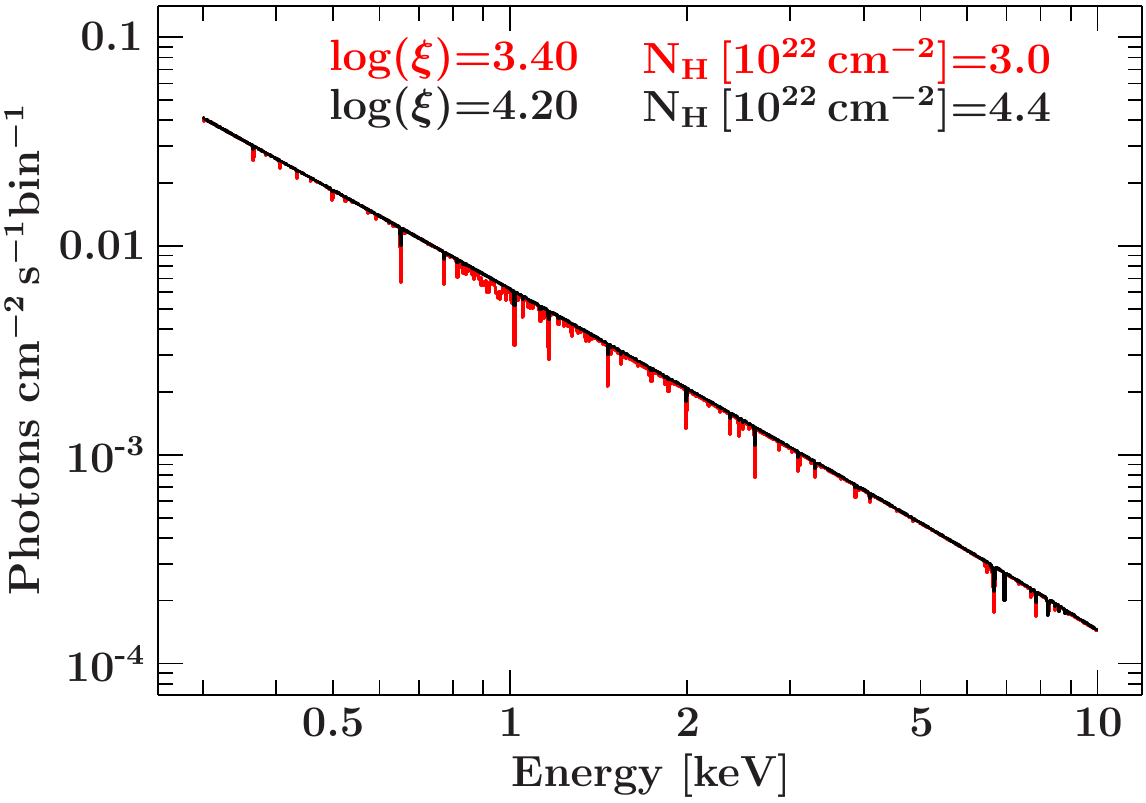}}
  \caption{Effect of \WA{3} if it were to fully cover the background
    power law. Shown in red is the warm absorber of observation Suz~1,
    black shows the effect of the absorber found in the remaining
    \suzaku observations.}
  \label{fig:zxipcf_high_logxi_nh}
\end{figure}
Figure~\ref{fig:zxipcf_high_logxi_nh} shows how both instances of
\WA{3} affect the high energy power law. The model with $\logxi \sim
3.4$ for Suz~1 imprints a clear absorption feature around
0.9--1.0\,keV due to \ion{Ne}{ix} and \ion{Ne}{x}. A column of
comparable ionization was also constrained by \citet{Markowitz2009}
but disappears in Suz~2 to Suz~6. It may, however, still be
intrinsically present but undetected because of degeneracies within
the model description. The highly ionized absorber \WA{3} ($\logxi\sim
4$) shows weak absorption lines from highly ionized species of ion
(Fig.~\ref{fig:zxipcf_high_logxi_nh}). These are difficult to
constrain in a single \suzaku observation but are consistent with the
detailed study of Suz~1--Suz~3 by \citet{Gofford2013}, who show the
existence of absorption lines due to He-like and H-like Fe.

\begin{figure}
  \resizebox{\hsize}{!}{\includegraphics{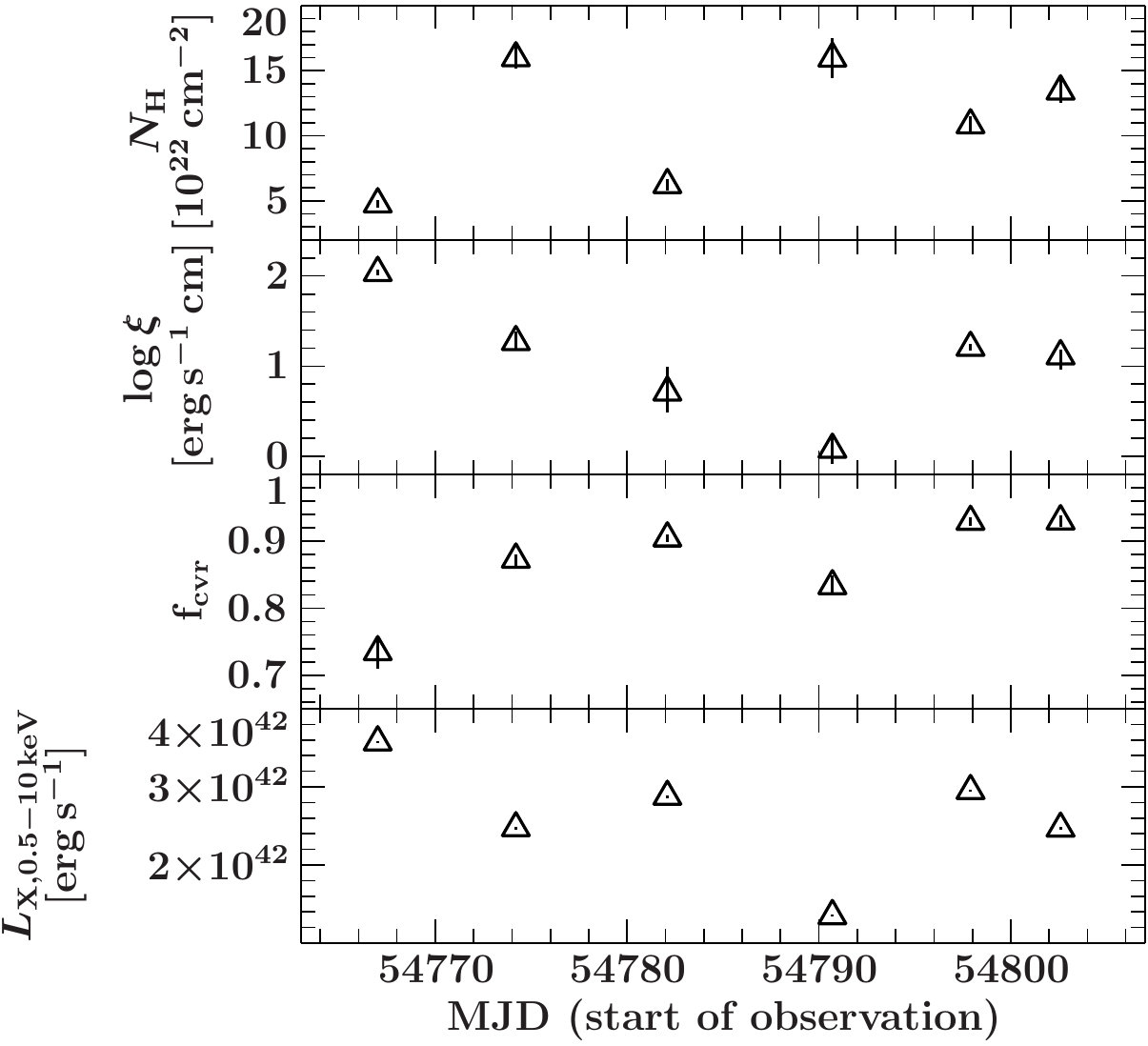}}
  \caption{Temporal evolution of the parameters describing the
    variable intermediately ionized absorber \WA{2} and the luminosity
    in the 0.6--10\,keV X-ray band for the six \suzaku observations.}
  \label{fig:par_lc}
\end{figure}

\begin{figure}
  \resizebox{\hsize}{!}{\includegraphics{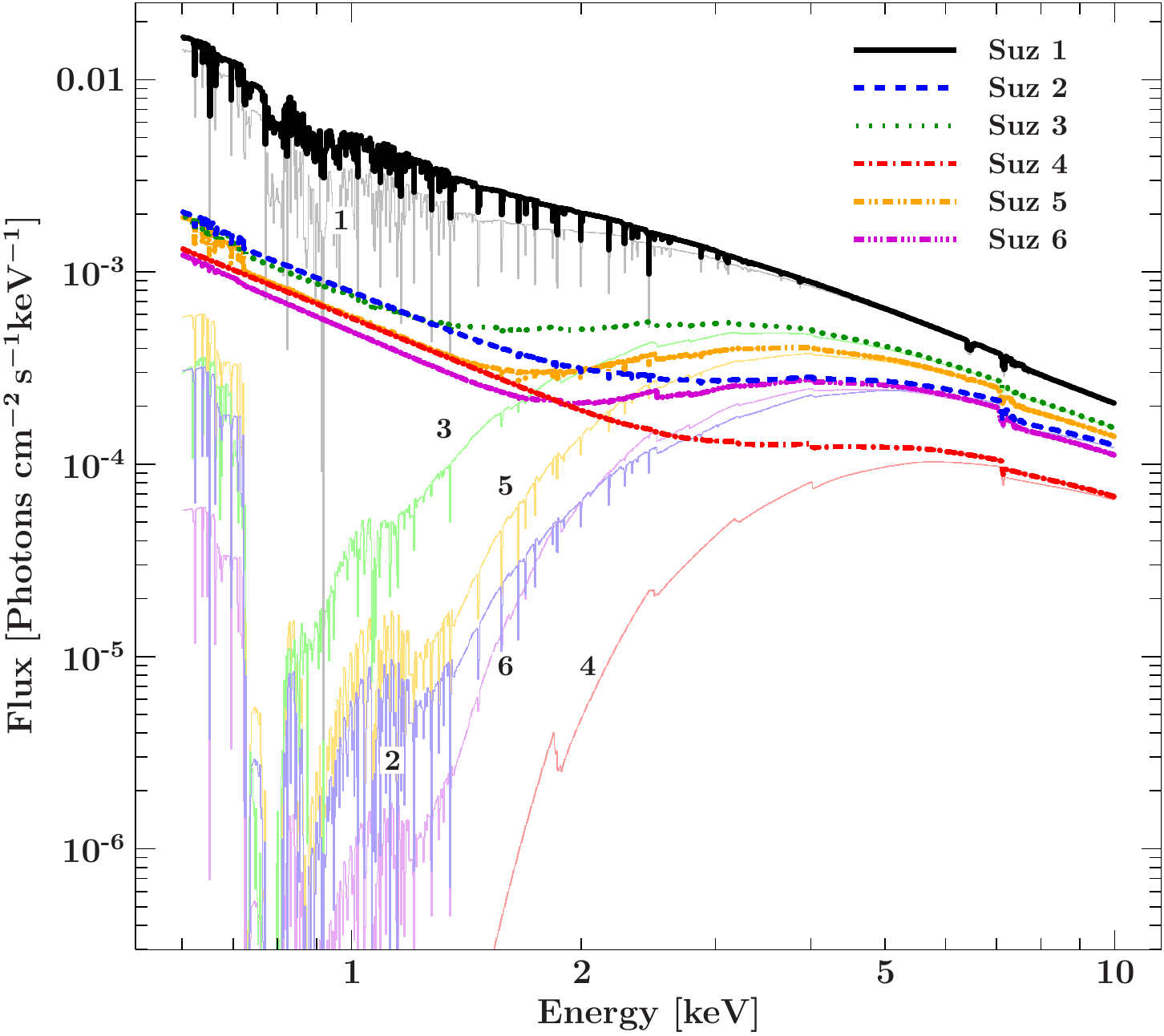}}
  \caption{Effect of the partial coverer \WA{2} onto a power law
    (thick lines with line styles given in the figure) as part of the
    best-fit model (Fig.~\ref{fig:all_suz_spectra}). The thin and
    solid lines of lighter color show the effect the coverer would
    have if it were to fully cover the source.}
  \label{fig:zxipcf_var_logxi_nh}
\end{figure}

\begin{figure}
  \resizebox{\hsize}{!}{\includegraphics{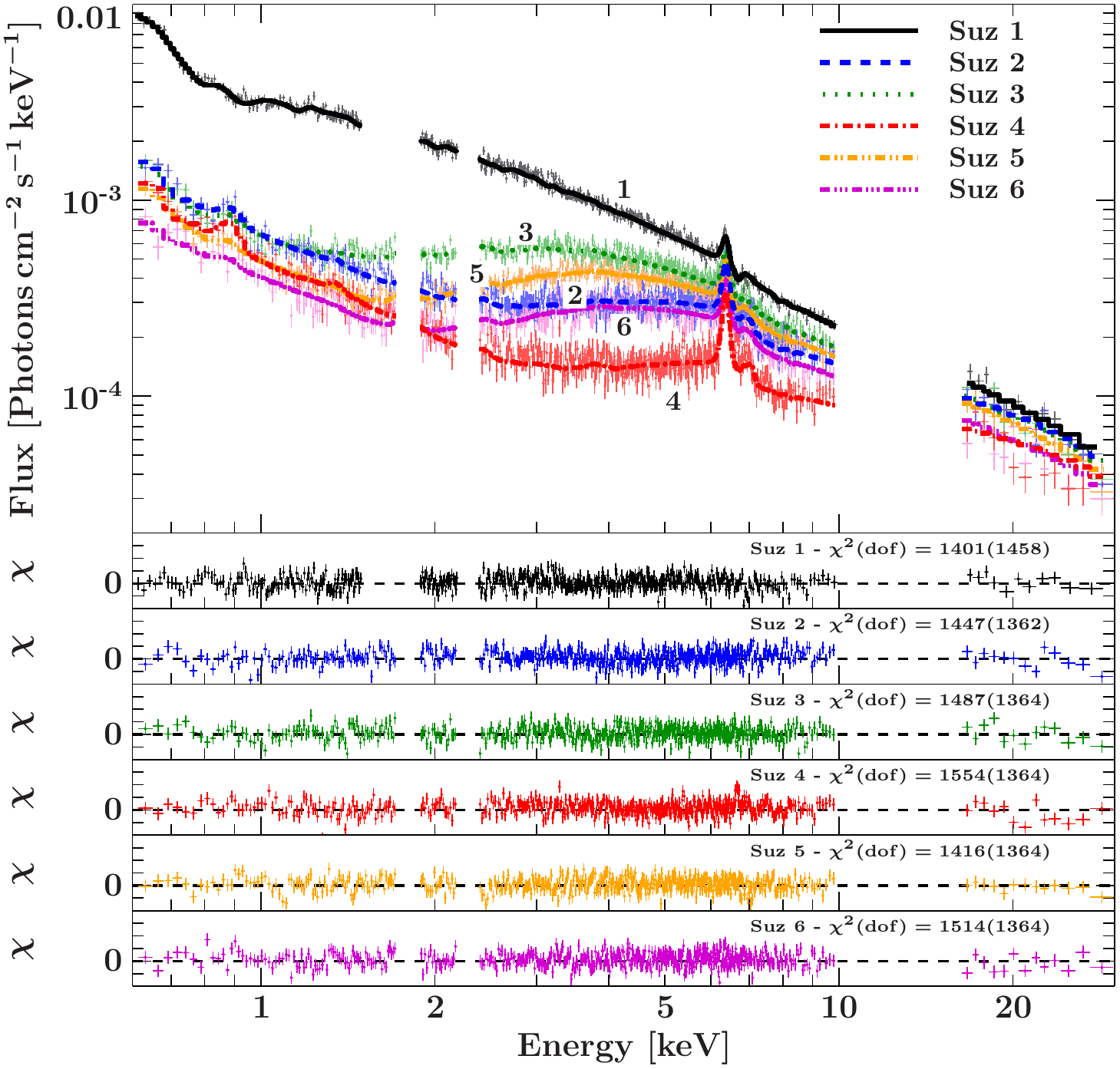}}
  \caption{Spectra and residuals in units of standard deviations for a
    fit of the baseline model to the \suzaku spectra of the
    2008 campaign (Table~\ref{tab:par_suz}). A single fit is performed
    for the high state observation (Suz~1) while the remaining
    observations (Suz~2--Suz~6) are fitted as part of one simultaneous
    fit with tied time-independent parameters. The statistics printed in the
    residuals panels correspond to the best-fit on each observation,
    using the appropriate time-dependent parameters.}
  \label{fig:all_suz_spectra}
\end{figure}

The \textit{time-dependent group parameters} of the model represent
the effects of variability. In particular we mention the variability
of the \textit{intermediately ionized partially covering absorber
  \WA{2}}. Simultaneous fitting isolates \WA{2} as the only variable
layer of absorption within the complex interaction of the three
absorbers contained in the model. Figure~\ref{fig:par_lc} shows the
time evolution of the fit parameters \nh, \logxi, $f_\mathrm{cvr}$ and
the X-ray luminosity $L_\mathrm{X}$. The covering fraction scatters
around $\sim$0.9. Only Suz~1 shows a lower value of $\sim$0.74. The
column density peaks in a complex fashion between
5--18$\times 10^{22}\,\mathrm{cm}^{22}$ across all six
observations.  The ionization parameter first decreases from $\logxi
\sim 2$ to nearly zero and then increases again to values of $\logxi
\sim 1$ for the last two observations.  In
Fig.~\ref{fig:zxipcf_var_logxi_nh} we illustrate the effect of \WA{2}
onto the power-law continuum if this absorber had a covering fraction
of unity (see also Fig.~\ref{fig:all_suz_spectra}). This
representation of the model helps to identify the most prominent
contributing absorption features below 1\,keV, which are mainly due
the He-like and H-like ions \ion{O}{vii}, \ion{O}{viii}, \ion{Ne}{ix},
and \ion{Ne}{x}\footnote{Note that
because in reality the warm absorber is only partly covering the
source, the fit is not particularly sensitive to these narrow features
and mainly driven by the characteristic curvature below 2\,keV caused
by the intermediate ionization of \WA{2}.}.

Other variable parameters are the photon-index $\Gamma$ that scatters
between $\sim$1.6 and $\sim$1.7.  We also find intermediate ionization
states of the reflecting and absorbing material ranging from
$\logxi\sim 0.8$ up to $\sim$2.

The results of the fits are shown in Figure~\ref{fig:all_suz_spectra}.
We show the XIS and HXD spectra of all observations including the
best-fit models. The residual panels suggest a fit that describes the
continuum well with adequate statistics. Figure~\ref{fig:all_suz_comp}
in the appendix gives an overview of all \suzaku spectra shown
together with the full model and the best-fit model
components. Clearly visible is an emission excess above the power law
in the energy range covered by \suzaku/HXD/PIN, most likely due to the
Compton hump peaking around 30\,keV. The 6.4\,keV Fe K$\alpha$ line is
self-consistently modeled by reflection off Compton-thick gas
(\texttt{xillver}), as is the extra emission below
1\,keV. Figure~\ref{fig:all_suz_comp} illustrates that the reflection
parameters are mainly derived from that soft emission and the iron
line complex as dominant features. The HXD/PIN data quality is good
enough to constrain the normalization of \texttt{xillver}. No extra
soft emission component is needed in contrast to the 2006 \xmm\
observation. Although we can not rule out its presence, the soft X-rays
are well modeled by a combination of ionized reflection and leaked
coronal emission described by the HXPL.  The (narrow) Fe~K$\alpha$
line has an equivalent width of $\mathrm{EW} \sim 130\,\mathrm{eV}$
and is well described by \texttt{xillver}. No further line emission is
required, indicating that the line is consistent with being completely
due to reflection off Compton-thick material.

\begin{figure}
  \resizebox{\hsize}{!}{\includegraphics{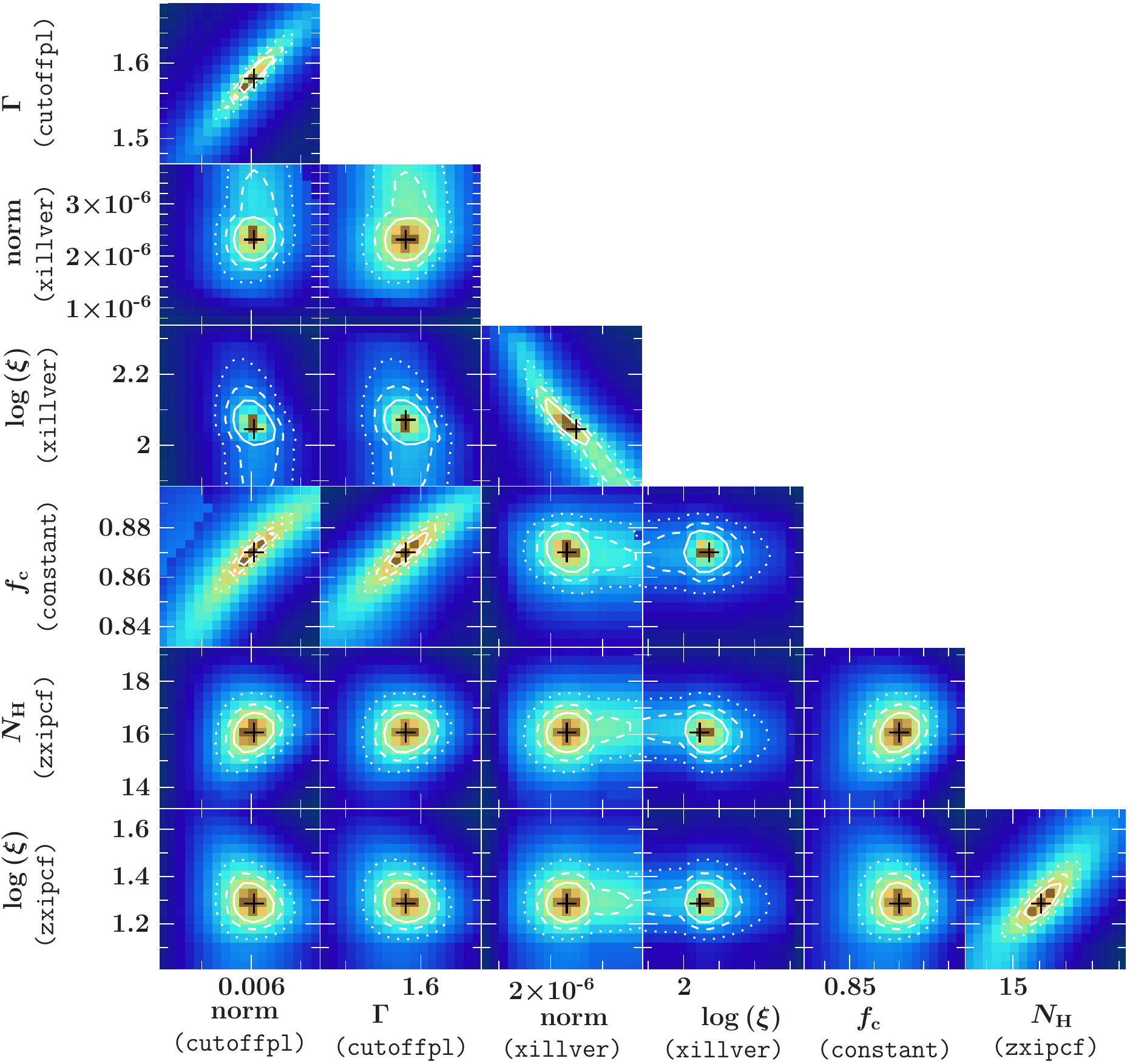}}
  \caption{Color coded $\Delta\chi^2$ contours for relevant spectral
    parameters of Suz~2. The 68.27\%, 90\% and 99\% contours
    correspond to the solid, dashed and dotted lines. The color code
    ranges from brown (small $\Delta\chi^2$) up to dark blue (large
    $\Delta\chi^2$).}
  \label{fig:cont_703022020}
\end{figure}

\begin{figure*}
  \centering
  \resizebox{0.9\hsize}{!}{\includegraphics{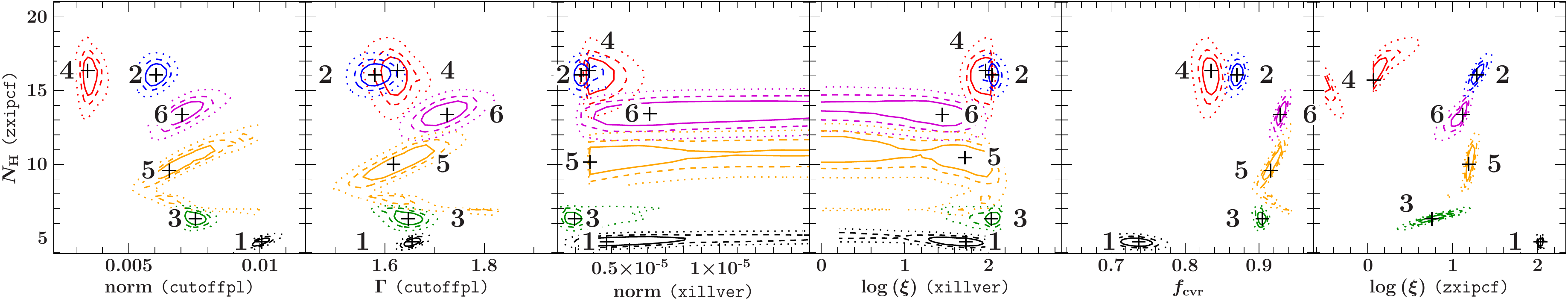}}
  \caption{Contours relating all time-dependent parameters of all
    observations to $N_\mathrm{H}$ of \WA{2}. The solid, dashed and
    dotted lines represent the 68\%, 90\%, and 99\% confidence
    contours. Observations Suz~1 to Suz~6 are both marked with numbers
    and color-coded in black, blue, green, red, orange, and purple,
    respectively.}
  \label{fig:nh_conf}
\end{figure*}

The complexity of our model leads us to study possible degeneracies by
calculating $\Delta\chi^2$ contours for all pairs of time-dependent
parameters. These contours show that the simultaneous fit is
remarkably robust. As an example, Fig.~\ref{fig:cont_703022020} shows
the contours for observation Suz~2. The overall correlations are
similar for the other observations but differ in extent depending on
the data S/N. A mild degeneracy between the covering factor with both
the normalization and the photon-index of the high energy power law is
present, as are modest degeneracies between the ionization state and
the column density of the \WA{2} as well as the photon-index and the
normalization of the power law. The contours of \texttt{xillver} show
an overall degeneracy with most other parameters, which can be removed
by assuming that photon-index is a time-independent
parameter. Finally, turning to the time variability of the absorber,
Fig.~\ref{fig:nh_conf} shows the contours of all time-dependent
parameters related to $N_\mathrm{H}$ for all observations and in
particular reveals a clear variability in column density across the
observations.  Note that the contours including the normalization and
ionization of the unblurred reflection component \texttt{xillver}
reveal no significant variability.

\subsection{Analysis of \swift\ XRT data}
\label{sec:analysis_swift}
For each \suzaku observation the archive contains two simultaneous
\swift\ observations. 

We model 2008 \swift\ XRT simultaneously with the same baseline
spectral model but independently of the simultaneous fit of the
\suzaku data. Tests show that the \swift\ data are consistent with
the \suzaku data and do not contribute with additional information
due to the lower effective area.  We therefore impose constraints
gained with \suzaku for certain parameters that can neither be
constrained with \swift\ nor identified as variable with \suzaku. All
derived parameters are consistent with those found for \suzaku data.

We also analyze the most recent \swift\ observations from
2013--2015. The data are well described with a single power law
absorbed by both Galactic and intrinsic, neutral absorption modeled
with \textsc{tbnew}. We find a very low degree of absorption
($N_\mathrm{H} < 0.34\times 10^{22}\,\mathrm{cm}^{-1}$) consistent
with the predominant relatively unabsorbed state during the \rxte\
monitoring over 6.9\,years.
\begin{figure}
  \centering
  \resizebox{\hsize}{!}{\includegraphics{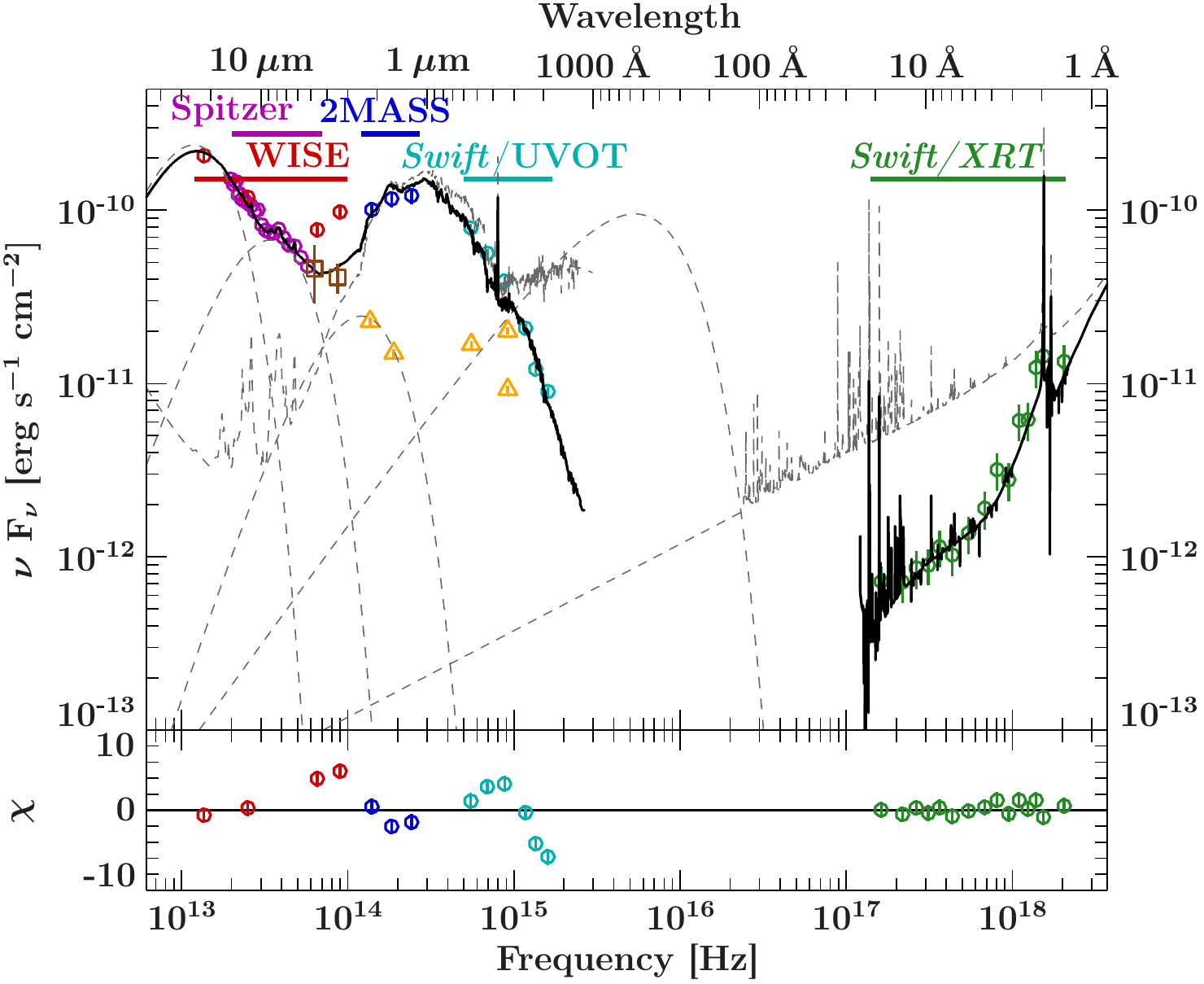}} \caption{Example-SED
    of NGC~3227 illustrating the coverage from the MIR (WISE: red
    circles; \spitzer: purple circles and \hst: orange triangles) over
    the NIR (ground-based data: brown squares), the optical (2MASS:
    blue circles; \hst) and UV (\swift/UVOT: turquoise circles) to the
    X-ray band up to 10\,keV (\swift/XRT: green circles). The UV and
    X-ray data were simultaneously measured at 2008-11-21 while the
    2MASS and WISE data are archival catalog data. The \spitzer data
    are simulated data based on the AGN contribution
    \citep{Caballero2015}. Other data are non-simultaneous photometric
    measurements from the literature. See text for details. The dashed
    lines correspond to the unabsorbed and unreddened model
    components, the thick solid line to the overall reddened and
    absorbed model.} \label{fig:sed}
\end{figure}

\section{Multiwavelength Data Analysis}

\subsection{NIR--X-ray SED and reddening}
\label{sec:spec_anal_reddening}
The broadband SED (Fig.~\ref{fig:sed}) is based on \swift/UVOT data
from the observation Sw~4a and allows us to estimate the amount of
reddening due to dust in the line of sight and to draw conclusions on
the dust content of the X-ray absorbing cloud. Model degeneracies,
however, require us to also include data from other wavebands. Those
are photometric data from ground- and space-based facilities of
different spatial resolutions and apertures: WISE \citep{Wright2010}
and 2MASS \citep{Skrutskie2006} data from the All-Sky Source Catalogs,
data \citep{RamosAlmeida2011} from the NASA 3\,m IRTF telescope
(NSFCam at L-band), the 3.8\,m UKIRT telescope (IRCAM3 at M-band), the
Gemini-South telescope (T-ReCs, N-band) as well as \hst NICMOS data in
the filters F160W and F222M \citep{Quillen1999}, \hst FGS data at
F550W \citep{Bentz2009} and \hst ACS data at 3300\AA\
\citep{Munoz2007}. This last data point, taken in 2002, is roughly
consistent with the UVOT flux taking into account the different
spatial resolution as well as likely long-term source
variability\footnote{The extended flag in the ALLWISE catalog
  introduces problems with the profile-fit photometry, which is why we
  add 10\% systematic uncertainties to 2MASS and WISE data as well as
  5\% for UVOT, which has a similar PSF.}. In addition, we simulate
\spitzer data based on the AGN contribution to the IR, as modeled by
\citet{Caballero2015}. A cross-calibration factor between WISE and
\spitzer of 1.6 is consistent with flux calibration uncertainties and
aperture effects \citep{Caballero2015}. For modeling the continuum, we
only take into account data with PSFs that are comparable to the one
of \swift/UVOT, i.e., WISE, 2MASS data with PSFs of
5\arcsec--10\arcsec radius, as well as \swift/XRT data. Systematic
effects and model degeneracies due to aperture effects of the
instruments and a lack of simultaneous data lead us to abstain from
performing a statistical $\chi^{2}$ minimization at energies lower
than the X-rays. Instead, we use sub-arcsec \spitzer, \hst and the
ground-based data that are only sensitive to the AGN core to constrain
the AGN-related model components, i.e., the dusty torus and the
accretion disk emission in the IR and UV, respectively. While the
Swift data provide the only variability information on timescales of
weeks, the X-ray band turns out to be the only band with strong
variability and we can use the already known model between
0.6--10\,keV. The IR--UV bands, in contrast, show only minor
variability up to 25\% on the time-scales monitored by \swift/UVOT. In
the following we describe our best fitting model for data below the
X-rays, for which we find $\chi^{2}/\mathrm{dof}=125/23$.

The reprocessed MIR/NIR emission probed by \spitzer, ground-based and
\hst data at 1550\AA\ and 2300\AA\ have been fitted previously with
clumpy torus models by, e.g., \citet{RamosAlmeida2011} or
\citet{AlonsoHerrero2011}. We instead use a phenomenological
description of three black-bodies from 145\,K up to 1480\,K that is
supported by \citet{Calderone2012}.

We take account of the host galaxy starlight using a SWIRE
\footnote{\url{http://www.iasf-milano.inaf.it/~polletta/templates/swire_templates.html}}
template for a Sa spiral galaxy. The broad-band template is
constructed to match a stellar template spectrum (3500--7000\AA) by
\citet{Winge1995} at 5500\AA\ for a normalization of one.  Below
4000\AA, UVOT measures extra emission that can not be fitted with an
accretion disk. According to \citet{Davies2006}, the central 30
parsecs show remainders of starburst emission accounting for 20\% to
60\% of the galaxy's bolometric luminosity. We therefore add a
starburst template \citep{Kinney1996} below 4000\AA. This is also
confirmed by an independent photometric measurement of the starlight
by \hst at 3300\AA\ (lower triangle in Fig.~\ref{fig:sed}) that
perfectly matches the combined template for a normalization of one. We
find a template normalization of $\sim$4, which is due to
uncertainties in the derivation of the stellar template and aperture
effects.

The accretion disk emission is described with a multi-temperature
disk-blackbody (\texttt{diskpn}, $T_\mathrm{max}\sim 1\times
10^{5}\,\mathrm{K}$ at $r_\mathrm{in}=6\,r_\mathrm{g}$ reddened with a
column of $1.2\times 10^{21}\,\mathrm{cm}^{-2}$ or $A_\mathrm{V} \sim
0.45\,\mathrm{mag}$). Its normalization is constrained by \hst data at
$5470\AA$ and $3300\AA$. The data have been taken 2 years apart from
each other, which can explain their scatter around the model.
\citet{Koratkar1999} confirm that the value found for $T_\mathrm{max}$
is consistent with sample studies of quasars with minimal extinction
as well as the expected inner disk temperature of a thin
$\alpha$-disk. Degeneracies between the applied reddening and the
inner disk temperature are reduced due to the consistency of the
reddening column with results from the literature. According to
\citet{Koratkar1999}, one usually assumes extinctions of
E(B-V)$\sim$0.05--1\,mag, ($N_\mathrm{H}\sim 0.04\times
10^{22}-0.8\times 10^{22}\,\mathrm{cm}^{-2}$ for a Galactic
gas-to-dust ratio). According to
\citet{Komossa1997,Kraemer2000,Crenshaw2001}, NGC~3227 has shown a
rather large reddening column of $\sim 0.2\times
10^{22}\,\mathrm{cm}^{-2}$ that is claimed to be consistent with a
persistent, dusty, ``lukewarm'' absorber at $\sim 100\,\mathrm{pc}$
distance. \swift\ observations from 2013--2015 likely measure the same
distant absorber but with only half the column, which is consistent
with the reddening found in our model and maybe also by
\citet{Winge1995}.

Compared to the variable X-ray-absorbing column densities measured
with \suzaku and \swift, the amount of UV extinction is associated
with a column that is $\sim 70$ times smaller.  Even for larger
reddening and therefore larger inner disk temperatures, the severe
mismatch to the X-ray column persists.  This is unexpected if both the
X-ray and UV-absorber would be of the same origin.

For the extinction at optical wavelengths
\citep[$A_\mathrm{V}$;][]{Fitzpatrick1999} we use the reddening-curve
$c(\lambda)$ from \citet{Crenshaw2001} that attenuates the intrinsic
flux $F_\mathrm{int}$ according to
\begin{equation}\label{eq:fmunred}
  F_\mathrm{obs} = F_\mathrm{int}\cdot 10^{-0.4 A_\lambda}
\end{equation}
with the extinction coefficient at the wavelength $\lambda$,
$A_\lambda=c(\lambda)\,E(\mathrm{B}-\mathrm{V})=c(\lambda)\,
A_\mathrm{V}/R_\mathrm{V}$. We choose $R_\mathrm{V}=3.1$ and estimate
the equivalent column density via the assumed Galactic gas-to-dust
ratio $N_\mathrm{H}=A_\mathrm{V}\cdot 2.69\times
10^{21}\,\mathrm{cm}^{-2}\,\mathrm{mag}^{-1}$ \citep{Nowak2012}. Note
that the standard Galactic reddening curve does not have to
apply for the intrinsic absorber in NGC~3227 where dust grains might
have a different composition and size distribution compared to the
Galaxy \citep{Crenshaw2001}.

\subsection{Optical Polarimetry}
\label{sec:opt_pol}
A portion of the optical AGN emission undergoes scattering into our
line of sight resulting in a few percent of linear
polarization. Investigations prefer the scenario of polar-scattering
for NGC~3227 \citep{Smith2004}. This result implies an
intermediate inclination of the nucleus towards the line of sight,
similar to what is predicted by clumpy torus models that are tested
for NGC~3227 in Sect.~\ref{sec:clumpytorus}. We therefore investigate
additional optical polarimetric data coincident to the 2008
\swift/\suzaku campaign that have been taken between 2008 October 26
and 2008 December 03 with the 84\,cm and 1.5\,m telescopes of the San
Pedro M\'artir observatory (SPM) in Mexico. The data were taken with
the B filter (Johnson system) as well as R and I filters (Cousins
system) according to the standard photometric system
\citep{Bessel2005}. For the magnitude-flux conversion we use the
zero-point magnitudes from \citet{Bessel1998}.
\begin{figure}
  \centering
  \resizebox{\columnwidth}{!}{\includegraphics{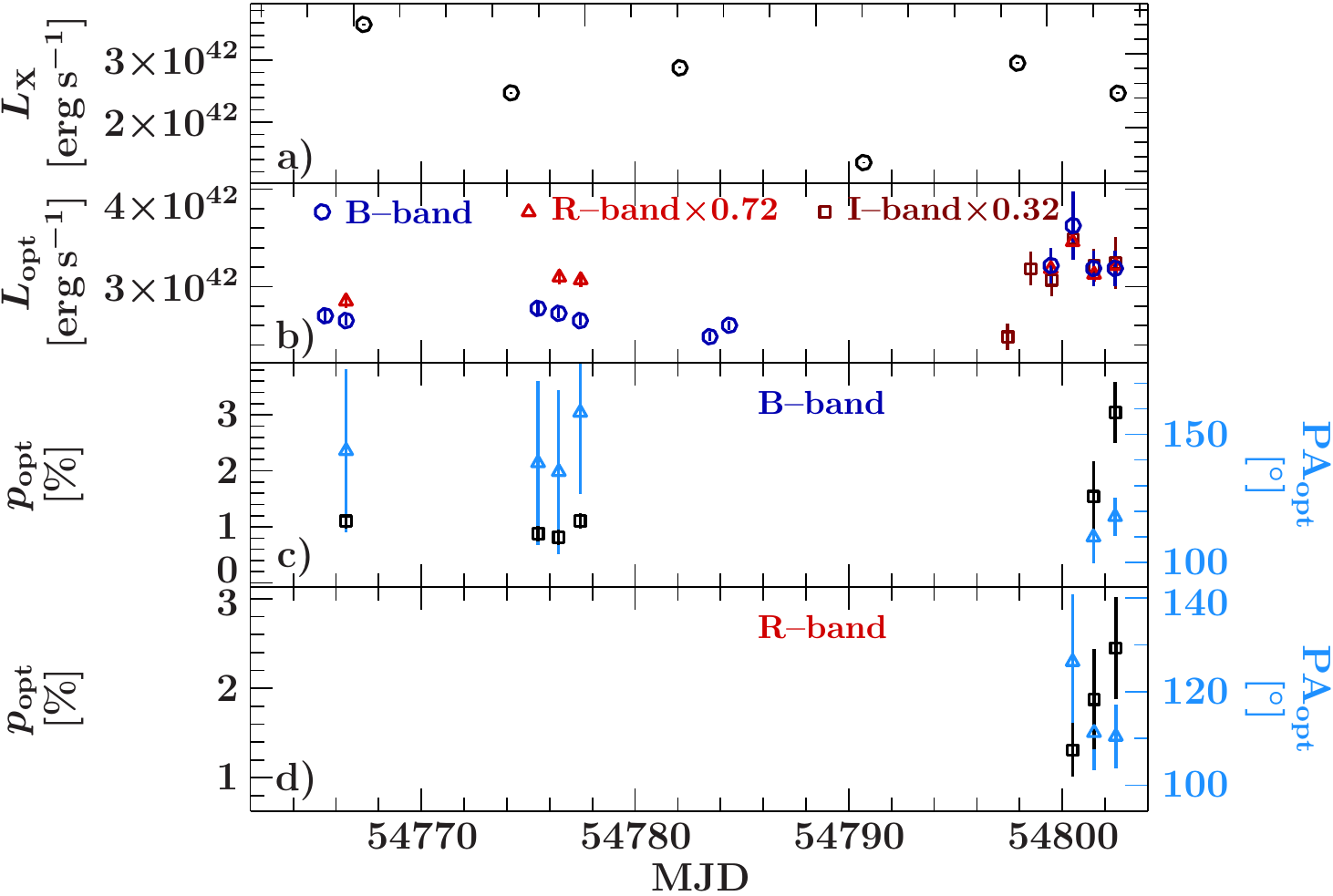}}
  \caption{\textbf{a} Light-curve of the unabsorbed X-ray luminosity
    in the $0.6$--$10\,\mathrm{keV}$ band; \textbf{b} light-curve of
    monochromatic luminosities in the B-band (blue circles), R-band
    (red triangles) and I-band (brown squares) for the 84\,cm and
    1.5\,m SPM telescopes; the R-band and I-band luminosities are
    normalized to the weighted mean of the B-band with relative
    factors of 0.72 and 0.32, respectively; \textbf{c/d} degree of
    optical polarization $p_\mathrm{opt}$ (black squares) and electric
    vector position angle ($\mathrm{PA}_\mathrm{opt}$, blue triangles)
    for the B/R-filters. The I-filter did not provide polarimetry
    data.  The central wavelength of the B/R/I-filters are given as
    $0.438/0.641/0.798\,\upmu\mathrm{m}$ by \citet{Bessel1998}.}
  \label{fig:lc_opt}
\end{figure}

We compare the optical photometric luminosities with the X-ray
luminosities in Fig.~\ref{fig:lc_opt}. The relatively sparse sampling
pattern precludes a detailed cross-correlation analysis between the
X-ray and optical light curves. The two bottom panels show polarimetry
data for the B- and R-band. The degree of polarization $p$ and the
electric vector position angle (PA) scatter around a few percent and
$100\degr$--$150\degr$, respectively, which is roughly consistent with
the results from \citet{Smith2004}. They find that the PA is
constant over wavelength with a value of $\mathrm{PA}\sim
135\degr$. In contrast we observe different weighted averages between
the B-band ($\mathrm{PA}\sim 140\degr$) and R-band ($\mathrm{PA}\sim
119\degr$). \citet{Mundell1995a} find both signatures for a narrow
line region in high-excitation \ion{O}{iii} emission lines
($\mathrm{PA}\sim 30\degr$) and two compact radio cores along a PA of
$\sim$170$\degr$. \citet{Mundell1995b} identify a much more extended
radio source with $\mathrm{PA}\sim 158\degr$ that shows a rotating
disk-like structure on scales of a typical host galaxy with an
inclination of $\sim$56$\degr$; any connection to the AGN on much
smaller scales is not immediately obvious, especially given the
uncertainty regarding the AGN system inclination. We support the
suggestion of \citet{Mundell1995a} and \citet{Smith2004} that the cone
of excited \ion{O}{iii} gas may be oriented along the symmetry axis of
the AGN. The angle enclosed between the cone and the measured average
PA in the B-band is therefore $\sim$102$\degr$, which is close to a PA
of $90\degr$. Together with the degree of polarization we therefore
find the optical emission in NGC~3227 to be consistent with
polar-scattering, similar to what is suggested by \citet{Smith2004}.

When facing additional systematic uncertainties in the polarimetry
data we can not conclude any significant variability.  The optical
luminosities, in contrast, show less than 40\% variability with
systematics most likely affecting the I-band data point at
MJD~54796.7.

\section{The properties and origin of the absorbing gas}\label{sec:origin}

\subsection{Summary of observational results}
\begin{figure}
  \centering
  \resizebox{0.8\hsize}{!}{\includegraphics{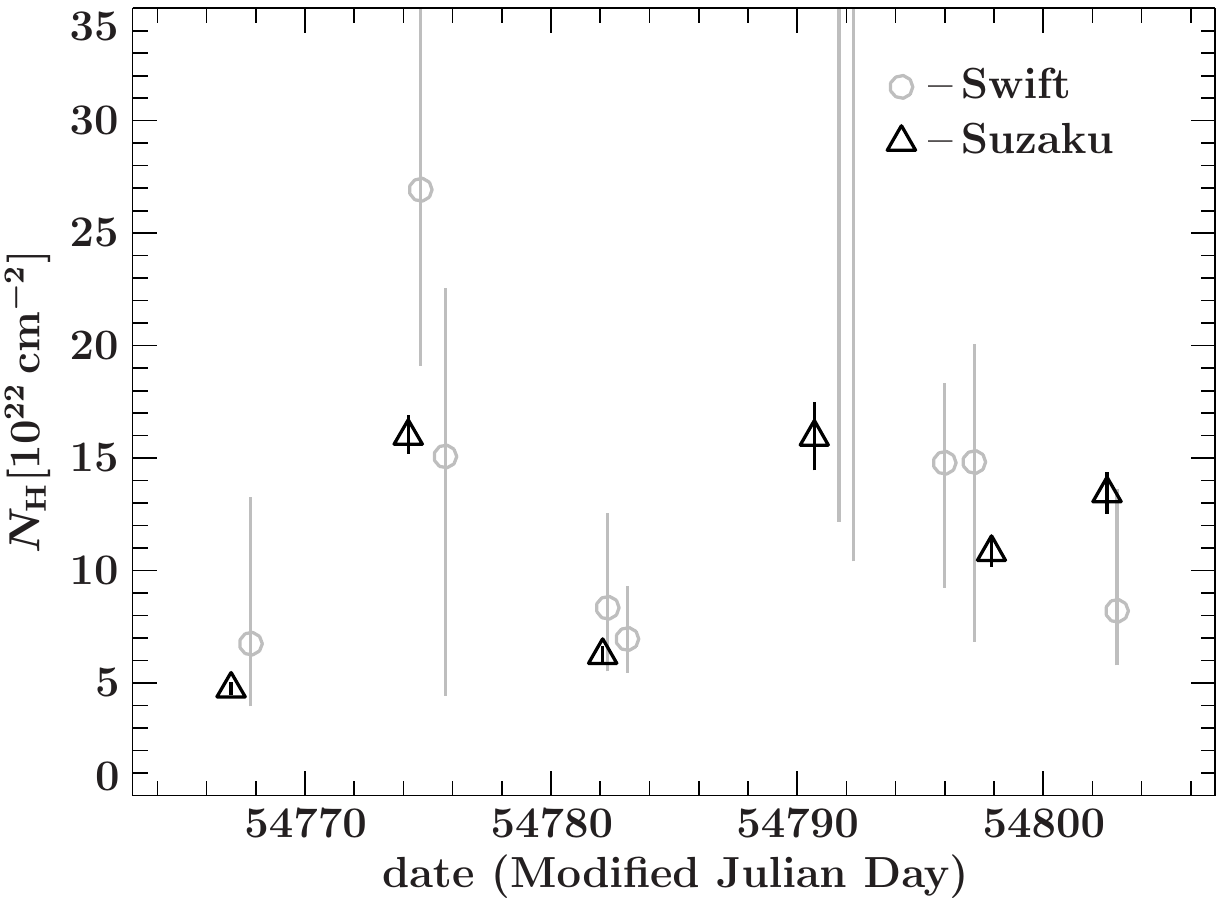}}
  \caption{Evolution of the column density $N_\mathrm{H}$ of \WA{2}
    over time as found in the \suzaku (black triangles) and \swift\
    (gray circles) data. We plot uncertainties according to the 90\%
    confidence level. The two data points derived by \swift\ that are
    cut off by the plotting window have the values $N_\mathrm{H} =
    42^{+22}_{-30}\times 10^{22}\mathrm{cm}^{-2}$ and
    $54^{+45}_{-38}\times 10^{22}\mathrm{cm}^{-2}$.}
  \label{fig:nh_lc}
\end{figure}

The 6.9\,years of sustained \rxte monitoring from 1999~Jan to
2005~Dec \citep{Rivers2011b} caught two discrete eclipses. The first
$\sim$80\,d long event in 2000 was analyzed by \citet{Lamer2003} and
is dominated by a lowly-ionized absorption component ($\logxi \sim
0.4$; \WA{1} in our model). The second event lasted 2--7\,days and was
also lowly- or at most moderately-ionized \citep{Markowitz2014}.
Based on this monitoring, the relatively unobscured observation by
\xmm in 2006,, and recent \swift\ observations from 2013--2015,
NGC~3227 was usually relatively unobscured before 2006 and after 2013.
Thanks to these previous observations and in particular the 2008
\suzaku/\swift\ campaign, we can build a clearer picture of the
behavior of the variable absorption components in NGC~3227 over the
last 15\,years and also address the properties and origin of the
absorber during the 2008 campaign.

During the 2008 campaign, the observed spectral variability was
dominated by absorption due to intermediately-ionized gas\footnote{We
  caution that \WA{2} as modeled during the 2008 campaign and \WA{2}
  as measured during the 2006 XMM observation may represent different
  components: the ionization parameters may be similar but the column
  densities differ by more than a factor of 50.} ($\logxi \sim 1$; \WA{2}
in our baseline model) with a time-variable, complex and irregular
density profile, and columns spanning $\sim$5--$16\times
10^{22}\,\mathrm{cm}^{-2}$ (Fig.~\ref{fig:nh_lc}). The column
densities derived by \swift\ are mostly consistent with those from
\suzaku within $2\sigma$. The event duration is $\geq
35\,\mathrm{days}$. Due to the lack of concurrent \rxte monitoring in
2008, we have no firm information on cloud ingress/egress.

\subsection{Luminosity and ionization}

\begin{figure}
\centering
  \resizebox{0.8\hsize}{!}{\includegraphics{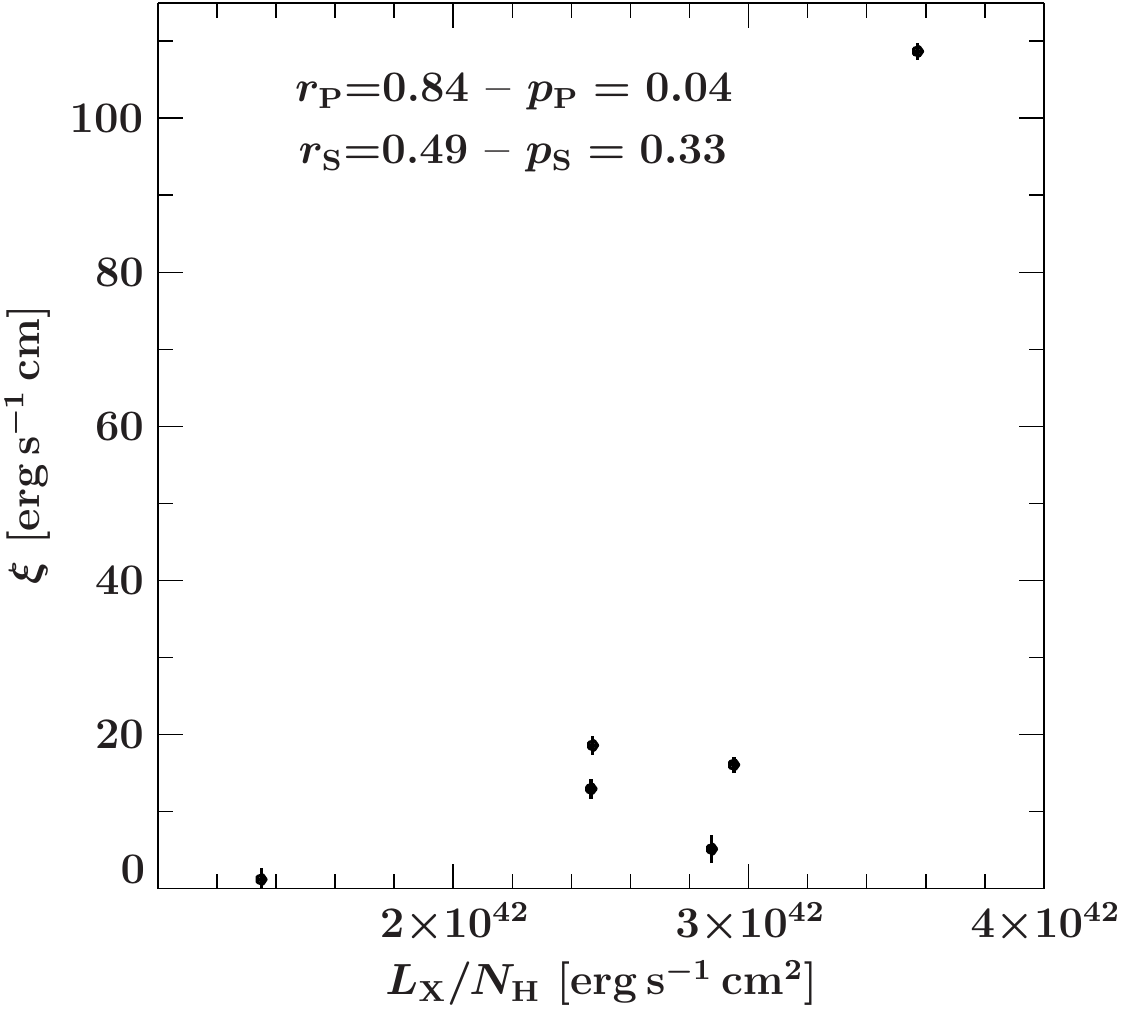}}
  \caption{Correlation between the ionization parameter $\xi$ and the
    X-ray luminosity $L_\mathrm{X}$ as a proxy for the ionizing
    luminosity. The numbers give the Pearson and Spearman rank
    correlation coefficients as well as the null hypothesis
    probabilities.}
  \label{fig:logxi-lion}
\end{figure}
We now proceed to study further the nature and properties of \WA{2} by
testing if there is any relation between its ionization state and the
measured source properties.
 
Because of Eq.~1, we expect a linear relation between
$\xi$ and the ionizing luminosity. We first assume that the measured
X-ray continuum is a direct proxy for the variation of the ionizing
continuum, i.e., we assume that the measured X-ray luminosity by
\suzaku, taken here from 0.6\,keV to 10\,keV, is directly
proportional to the overall ionizing luminosity
$L_\mathrm{ion}$. Figure~\ref{fig:logxi-lion} shows that there may be
a trend in the data that higher luminosities also imply larger
$\xi$. The Pearson correlation coefficient for $\xi$ as a function of
$L_\mathrm{X}$ is $r_\mathrm{P,X}=0.70$ with a null hypothesis
probability of $P_\mathrm{P,X}=0.12$. However, the Pearson correlation
coefficient is sensitive to extreme outliers, i.e., the data from
Suz~4. As a second check we also calculate the Spearman rank
correlation coefficient, which is less sensitive to extreme outlying
points than the Pearson coefficient. We find $r_\mathrm{S,X}=0.71$
($P_\mathrm{S,X}=0.11$). The data thus do not allow us to claim a
direct relation between $\xi$ and $L_\mathrm{X}$; any
correlation is tentative at best. More observations,
especially at lower luminosity states, are needed.

\subsection{The location of the absorber}
\label{sec:location}
Here we provide estimates on the distance of the variable absorber
detected in the 2008 campaign from the SMBH. Throughout this section
we make the simplified assumption that the mapped column density
profile is due to a single spherical cloud of uniform density,
illuminated by a central source and therefore ionized. For the
ionization state and column density we will use the average fit
parameters from the \suzaku campaign, i.e., $\log \xi\sim 1.1$ and
$N_\mathrm{H}\sim11.2\times 10^{22}\,\mathrm{cm}^{-2}$.

Many of the properties of the cloud also depend on the ionizing
luminosity, $L_\mathrm{ion}$. As we showed above, there is only a
tentative relation between the ionization parameter and the X-ray
luminosity as proxy for the ionizing luminosity, i.e., the ionization
state of the absorber is mainly determined by the source behavior in
the UV, which is inaccessible to our measurements and hidden in model
degeneracies. We can place a constraint on the source's UV emission
from the measured SED shown in Fig.~\ref{fig:sed}.

In order to derive the ionizing luminosity between 13.6\,eV and
13.6\,keV (1--1000\,Ryd), we assume that the UVOT data are non-variable
within the 2008 campaign.  We find $L_\mathrm{ion} = 8.9\times
10^{42}\,\mathrm{erg}\,\mathrm{s}^{-1}$ for the sum of the
non-variable disk-blackbody and the average of the X-ray power laws of
all observations. The average bolometric luminosity is measured as
$L_\mathrm{bol} = 1.3\times
10^{43}\,\mathrm{erg}\,\mathrm{s}^{-1}$. Using the assumed black hole
mass, we find an Eddington ratio of $\lambda_\mathrm{Edd}\sim 0.005$.

The measured value of $L_\mathrm{ion}$ is highly model dependent. We
therefore compare our results with those of \citet{Vasudevan2009} and
\citet{Vasudevan2010}, who present strong sample studies for
estimating the energy output of AGN but lack extended data coverage.
They show that the ratio of the UV-luminosity below 100\,eV to the
bolometric luminosity,
$L^\mathrm{ion}_{0.0136-0.1\,\mathrm{keV}}/L_\mathrm{Bol}= 0.21\ldots
0.59$ for values of the Eddington ratio
$\lambda_\mathrm{Edd}=L_\mathrm{Bol}/L_\mathrm{Edd}$ ranging from 0.01
to 0.61. In contrast to our study, \citet{Vasudevan2010} include
nuclear IR emission in the bolometric luminosity
$L_\mathrm{Bol}=10^{43.5}\,\mathrm{erg}\,\mathrm{s}^{-1}$ for
NGC~3227. Their lower black hole mass results in a higher Eddington
ratio of $\lambda_\mathrm{Edd}=0.039$. We find via linear
interpolation, that $L^\mathrm{ion}_{0.0136-0.1\,\mathrm{keV}} \sim
0.23\,L_\mathrm{Bol} \sim 10^{42.9}\,\mathrm{erg}\,\mathrm{s}^{-1}$,
which results in $L^\mathrm{ion}_{0.0136-13.6\,\mathrm{keV}}\sim
1\times 10^{43}\,\mathrm{erg}\,\mathrm{s}^{-1}$ using our measured
X-ray data. Considering the uncertainties of this method, this value
is consistent with our result from a more complete SED in a single
source study.

\subsubsection{Constraints from the ionization parameter}
\label{sec:ion_constr}
We assume that the illuminated cloud is in photoionization equilibrium
with the radiation field. Using the definition of $\xi$
(Eq.~1)
and estimating the particle density from the radial extent, $\Delta
R$ ($\le R$), of the cloud, $n_\mathrm{H}=N_\mathrm{H}/\Delta R$, we find
\begin{equation} \label{eq:r_ion}
 R\leq \frac{L_\mathrm{ion}}{\xi\,N_\mathrm{H}},
\end{equation}
This yields an upper limit for the distance of the cloud of 2.2\,pc
(2938\,ld) when using the average measured parameters.

\subsubsection{Constraints from the photo-ionization equilibrium}
\label{sec:phion_equilibrium_constr}

Following \citet{Reynolds1995}, we now combine the information we gain
from Eq.~3 with requirements for the recombination time
scale. As we assume that the absorber is in photo-ionization
equilibrium with the radiation field, the recombination timescale of
the plasma must be much smaller than the timescale of variations of
the ionizing continuum. A direct test of this assumption would be
measurements of a positive correlation between $\logxi$ and
$L_\mathrm{ion}$ or $L_\mathrm{X}$. Figure~\ref{fig:logxi-lion} indeed
shows a positive trend, but the correlation is weak.  This can be
explained with the assumption of a spherical cloud, which may be an
oversimplification. Any variations of the extend or number density of
the cloud can distort a direct correlation. Also, the ionizing
luminosity can not be measured directly, as the gap between the
far-UV and soft X-rays is not observable. If the cloud reaches
ionization equilibrium, the recombination time-scale is given by
\citep{Blustin2005,Krolik1999}
\begin{equation}\label{eq:t_rec}
  t_\mathrm{rec} \sim (n_\mathrm{e}\alpha_\mathrm{rec})^{-1}\sim 2\times 10^{4}Z^{-2}T_{5}^{1/2}n_\mathrm{9}^{-1}\,\mathrm{s}
\end{equation}
where $n_9=n_\mathrm{e}/10^9\,\mathrm{cm}^{-3}$ with $n_\mathrm{e}$
being the electron particle density, and where $\alpha_\mathrm{rec}$
is the recombination rate coefficient, $Z$ the atomic number of the
ion, and $T_{5}=T/10^5\,\mathrm{K}\sim 1$, a representative value
corresponding to gas with the ionization parameter $\logxi \gtrsim 1$
\citep{Reynolds1995,Krolik2001,Chakravorty2009}. We use $Z=9$ to
represent the likely dominant ions for \WA{2}, \ion{O}{viii} and
\ion{Ne}{ix} \citep{Kallman2001}. In their study of warm absorber
properties, \citet{Chakravorty2009} find that for typical AGN SEDs the
$\log\xi/T$-$\log T$ stability curves are independent of the hydrogen
number density for $10^{5} \leq n_\mathrm{H} \leq
10^{10}\,\mathrm{cm}^{-3}$. Assuming solar abundances, 
$n_\mathrm{e}=1.2 n_\mathrm{H}$. Using these assumptions we find
$t_\mathrm{rec} \sim 370\,n_\mathrm{9}^{-1}\,\mathrm{s}$, suggesting
recombination timescales easily less than hours--days for most
densities of relevance here.

\begin{figure}
  \resizebox{\hsize}{!}{\includegraphics{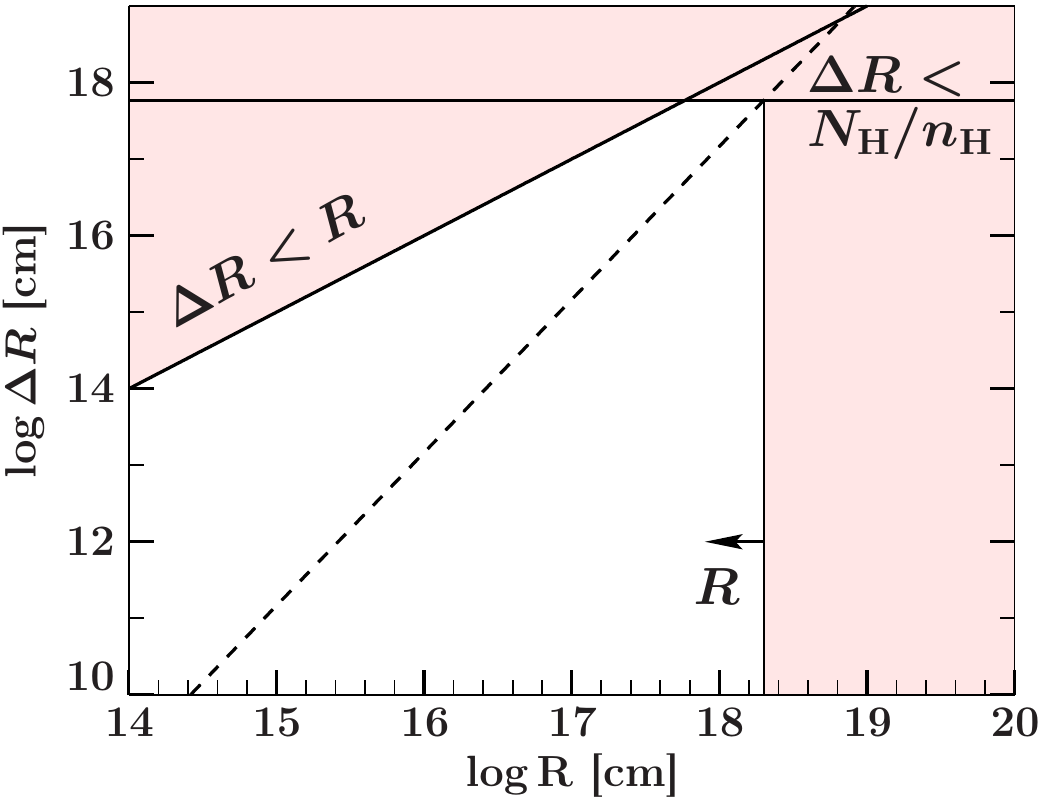}}
  \caption{Constraints on the radial distance $R$ of the ionized
    absorbing material from the central engine visualized in the $\log
    R - \log \Delta R$ plane, where $\Delta R$ resembles the diameter
    of the intrinsic absorber. The regions filled with light-red can
    be excluded based on spectral results of \suzaku and the
    relations denoted along the border lines. The dashed line follows
    the relation $\Delta R = N_\mathrm{H}R^{2}\xi/L_\mathrm{ion}$ with
    the appropriate average parameters derived from spectral fits to
    all \suzaku observations. The intersecting points with line where
    $\Delta R<N_\mathrm{H}/n_\mathrm{H}$ marks the upper limit on
    $R$.}
  \label{fig:distance_wa}
\end{figure}

A further constraint on the density can be obtained from the observed
variability of the absorber. The shortest timescales over which we can
reliably measure changes in the illuminating flux and ionization in
NGC~3227 is 7\,days. \rxte light-curves indicate that NGC~3227's X-ray
flux typically varies by factors of a few tens of percent on
timescales of 7\,days and less \citep{Uttley2005}, which makes the
assumption of photoionization balance a reasonable assumption. The
weekly spaced \suzaku observations confirm this finding, while no
variability is found within each of the observations except of
Suz~1. The overall low variability is consistent with the assumption
of photo ionization balance, such that $t_\mathrm{rec}\leq
t_\mathrm{var}=7\,\mathrm{d}$. This requirement can be translated to
find a lower-limit on the hydrogen number density from
Eq.~4,
\begin{equation}  \label{eq:n_H_low}
  n_\mathrm{H} > \frac{2\times 10^{4}}{1.2\,Z^{2}\,t_\mathrm{var}}\sim 1.94\times 10^{5}\,\mathrm{cm}^{-3}
\end{equation}
which, in turn, sets an upper limit on the thickness of the absorbing
layer,
\begin{equation}
  \label{eq:constr_dR}
  \Delta R < \frac{N_\mathrm{H}}{n_\mathrm{H}}.
\end{equation}
Figure~\ref{fig:distance_wa} helps to find an appropriate upper limit
on the distance $R$ of the absorber by considering the variation of
$\log \Delta R$ against $\log R$ \citep[see also][]{Reynolds1995}. We
can exclude three regions in the $\log R$-$\log \Delta R$ space
(Fig.~\ref{fig:distance_wa}, light-red region). The first region is
obtained by using Eq.~6 with the assumed average
$N_\mathrm{H}$ of the absorbing cloud and the minimum hydrogen number
density necessary to obtain ionization balance. In addition we can
also exclude a region where the basic assumption $\Delta R < R$ is
violated. Since
\begin{equation}\label{eq:dR-R}
  \Delta R = \frac{N_\mathrm{H}R^{2}\xi}{L_\mathrm{ion}}
\end{equation}
and because of $N_\mathrm{H}=n_\mathrm{H}\Delta R$ we can also find an
upper limit for $R$. Inserting the average $N_\mathrm{H}$, ionization
state, and luminosity measured with \suzaku gives the dashed line in
Fig.~\ref{fig:distance_wa}. 
An upper limit for $R$ can be derived from the intersecting point of
Eq.~6 and Eq.~7. As a result, the blank
region includes all values of $\Delta R$ and $R$ that come into
consideration based on our spectral analysis, i.e., $R \lesssim
10^{18.3}\,\mathrm{cm}=0.65\,\mathrm{pc}=770\,\mathrm{ld}$ (blue
upper limit in Fig.~\ref{fig:distance}), consistent with the dusty
torus following \citet{Suganuma2006}, \citet{RamosAlmeida2011} and our
geometrical considerations in Eq.~3.

\subsubsection{Constraints from a putative orbiting cloud}
\label{sec:orbit_constr}

In this section we additionally consider the Keplerian orbital motion
of an obscuring cloud that is illuminated by the central source while
passing the line of sight in $\geq$35\,d, equal to the duration of the
campaign.  We again use the average parameters $N_\mathrm{H}$, \logxi,
and $L_\mathrm{ion}$. With these assumptions we can estimate the
distance of such a cloud.

\begin{figure*}
  \centering
  \resizebox{0.7\hsize}{!}{\includegraphics{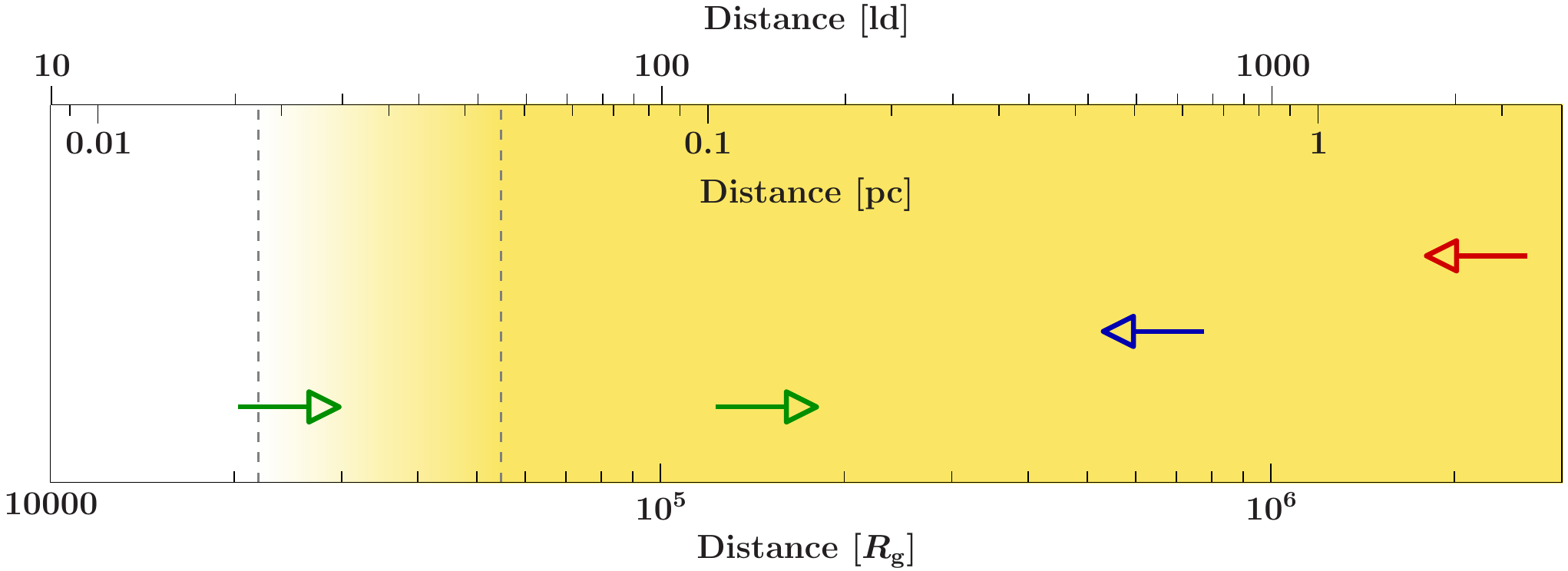}}
  \caption{Radial distribution of clumps that can potentially exist
    below and beyond the dust sublimation zone, i.e.,
    $0.4\,R_\mathrm{d}$--$R_\mathrm{d}$ (dashed lines). The distance
    limits from Sect.~\ref{sec:location} are marked with colored
    arrows. For all estimates we use the average measured parameters
    of \WA{2} plus the average ionizing luminosity. The red upper
    limit from Sect.~\ref{sec:ion_constr} uses the definition of the
    ionization parameter, while the blue upper limit from
    Sect.~\ref{sec:phion_equilibrium_constr} is determined from
    geometrical constraints on a spherical homogeneous cloud in
    photoionization balance. The Keplerian orbital motion of such a
    hypothetical cloud passing the line of sight is additionally
    included to form the lower-limit in green
    (Sect.~\ref{sec:orbit_constr}) while the second, larger
    lower-limit in green is determined for an orbiting cloud
    neglecting the information on its ionization or column density. }
  \label{fig:distance}
\end{figure*}

The first rough estimate is based on
\citet{Risaliti2007,Risaliti2009b,Risaliti2009a} and
\citet{Lohfink2012}, who discuss a spherical cloud that passes the
line of sight. It is able to fully cover the central X-ray source, if
it is larger than the diameter of the X-ray emitting corona
$\sim$10$\,r_\mathrm{g}$ \citep{Morgan2012} to five times this value,
which is arbitrarily chosen. This results in a lower limit for the
distance of such a cloud of $R \gtrsim 0.1\,\mathrm{pc}$, marked as
green lower limit in Fig.~\ref{fig:distance}. Due to the complex shape
of the $N_\mathrm{H}$ profile, the underlying assumptions are probably
too simple. The limit is also highly sensitive to the choice of the
size of the X-ray emitting region.

This very rough estimate of the distance can be significantly improved
when adding the information on the average irradiating luminosity as
well as column density and ionization state of the putative cloud.
Writing the Kepler velocity as $v=\sqrt{G\,M_\mathrm{BH}/R}=\Delta
R/\Delta t=N_\mathrm{H}/\left(n_\mathrm{H}\,\Delta t\right)$ and
expressing $n_\mathrm{H}$ in terms of the ionization parameter gives
\citep{Lamer2003},
\begin{equation} \label{eq:R_Lamer}
  R \geq 4\times
  10^{16}\,\mathrm{cm}\,M_{7}^{1/5}\left(\frac{L_\mathrm{ion}\,\Delta
    t_\mathrm{days}}{N_{22}\,\xi}\right)^{2/5} 
\end{equation}
With $M_7=M_\mathrm{BH}/10^7\,M_\odot=0.76$,
$L_{42}=L/10^{42}\,\mathrm{erg}\,\mathrm{s}^{-1}=8.9$,
$N_{22}=N_\mathrm{H}/10^{22}\,\mathrm{cm}^{-2}=11.18$, and $\Delta t
\geq 35$\,d we find $R\geq 0.017\,\mathrm{pc}=20.7\,\mathrm{ld}$. This
lower limit is shown as additional green lower limit in
Fig.~\ref{fig:distance}.

Finally, we estimate the size $\Delta R$ of the assumed spherical
cloud. Equating the Keplerian velocity with $\Delta R/\Delta
t=N_\mathrm{H}/n_\mathrm{H}\,\Delta t$, we find
\begin{equation}\label{eq:density}
  n_\mathrm{H}=\frac{N_\mathrm{H}}{\Delta t}\sqrt{\frac{R}{G\,M_\mathrm{BH}}}.
\end{equation}
Our distance estimates of 0.017--0.65\,pc and the campaign-average
value for $N_\mathrm{H}$, then yield number densities of
$n_\mathrm{H}\sim 2.7\times 10^{8}$--$1.7\times
10^{9}\,\mathrm{cm}^{-3}$, consistent with the lower-limit found from
the recombination time-scale analysis. This result translates to a
range in the size of the absorbers of $\Delta R \sim 6.6\times
10^{13}$--$4.1\times 10^{14}\,\mathrm{cm}$. Note that the column
density profile has no defined ingress or egress. If we assume that
the absorption by the cloud takes $\lesssim 2\,\mathrm{years}$, i.e.,
the time interval between the absorbed observations and the 2006
unabsorbed \xmm\ observation, the density would be about one order of
magnitude less and its size accordingly larger.

\subsection{A dust-free cloud located in the BLR?}
\label{sec:nondusty_cloud}
The range of 0.017--0.65 pc found in the previous section means the
possible location for the cloud spans radial distances both inside and
outside the dust sublimation zone. Here, we define the dust
sublimation zone to range from $0.4\,R_\mathrm{d}$--$R_\mathrm{d}$,
with the the dust sublimation radius,
\begin{equation}
  \label{eq:dustsubl}
  R_\mathrm{d}=0.4\,\mathrm{pc}\cdot
  \left(\frac{L_\mathrm{bol}}{10^{45}\,\mathrm{erg}\,\mathrm{s}^{-1}}\right)^{0.5}\left(\frac{T_\mathrm{d}}{1500\,\mathrm{K}}\right)^{-2.6}\sim 0.05\,\mathrm{pc}
\end{equation}
with an assumed bolometric luminosity of
$1.3\times 10^{43}\,\mathrm{erg}\,\mathrm{s}^{-1}$ and a
dust temperature of $T_\mathrm{d}=1500\,\mathrm{K}$
\citep{Barvainis1987,Nenkova2008b}. 

The result is consistent with \citet{Blustin2005}, who determine the
minimum and maximum distance of the absorbing gas solely based on
geometrical considerations for a sample of 23 Seyfert galaxies of
intermediate classification.  \citeauthor{Blustin2005} find that most
low-velocity absorbers are consistent with the inner edge of the
torus. Similarly, \citet{Risaliti2002} explain the column density
variations of a large fraction of their Seyfert~2 sample sources with
clouds at the inner edge of the dust sublimation zone.

We can test the dust content of the cloud by using the reddening
derived in Sect.~\ref{sec:spec_anal_reddening}. If the Galactic
dust-to-gas ratio is applicable, the X-ray absorbing gas columns
predict that we should see roughly $A_\mathrm{V} \sim 20$--60\,mag of
optical extinction. During the strong 2008 absorption event, however,
we only measure a reddening of $A_\mathrm{V} \sim
0.45\,\mathrm{mag}\sim 1.2 \times 10^{21}\,\mathrm{cm}^{-2}$. 

There are several potential reasons for the lack of a strong dust
component in the variable X-ray column density.  First we consider a
scenario where a cloud inside the line of sight to the X-ray source
may indeed contain dust, but does not cover the line of sight to the
UV continuum source. The diameter of a spherical cloud on a Keplerian
orbit was determined in Sect.~\ref{sec:orbit_constr} to be
$\sim$$6.6\times 10^{13}$--$4.1\times 10^{14}\,\mathrm{cm}$. We now
estimate the diameter of the UV emitting part of the accretion disk
for comparison.
The radial temperature profile of a standard thin disk is
\begin{equation}
  \label{eq:accrdisctemp}
  T(R)\sim 2\times
  10^{5}\,\mathrm{K}\,\left(\frac{M}{10^{8}\,M_{\astrosun}}\right)^{1/4}\,\left(\frac{\dot{M}}{\dot{M}_\mathrm{Edd}}\right)^{1/4}\,\left(\frac{R}{10^{14}\,\mathrm{cm}}\right)^{-3/4}
\end{equation}
with the derived Eddington ratio
$\lambda_\mathrm{Edd}=L_\mathrm{bol}/L_\mathrm{Edd}=\dot{M}/\dot{M}_\mathrm{Edd}\sim
0.004$, the black hole mass and the radius of the UV emitting region.
To solve the equation for $R$, we estimate the temperature at the
outer UV emitting disk using the Wien displacement law and the longest
wavelength \swift/UVOT filter. We find a radius of $1.1\times
10^{15}\,\mathrm{cm}$, i.e., larger than the estimated diameter range
of a spherical, homogeneous X-ray absorbing cloud.  In addition,
\citet{McHardy2014}, studying the interband time lags in \object{NGC~5548},
inferred that UV-emitting regions can extend slightly further than
expected from standard $\alpha$-disk theory, at least in that object.
Consequently, based on our results that assume a simple spherical
cloud, we cannot rule out that the weak reddening is due to a X-ray
absorbing cloud that fully covers the X-ray corona but not the entire
UV-emitting disk.

A second possible reason for the low reddening is that the cloud does
indeed contain dust but at an extremely high gas-to-dust ratio. With
the results from above, we find a gas-to-dust ratio of
$N_\mathrm{H}/A_\mathrm{V}\sim 2.5\times
10^{23}\,\mathrm{cm}^{-2}\,\mathrm{mag}^{-1}$, which is about a factor
100 higher than the assumed Galactic gas-to-dust ratio.

Finally, the low level of variability that we infer for the reddening
supports the notion that the reddening matter has an origin that is
distinct from the much more strongly variable X-ray column. It is
consistent with the distant, dusty, ``lukewarm'' absorber identified
by \citet{Kraemer2000} and \citet{Crenshaw2001} as well as recent
\swift\ observations.  They therefore conclude that the X-ray
absorbing gas is likely dust-free, which is supported by
Sect.~\ref{sec:orbit_constr}. We find that the absorber can be located
well below the dust sublimation radius, and thus plausibly be a
BLR-cloud. The hydrogen number density range of $2.5\times
10^{8}$--$1.5\times 10^{9}\,\mathrm{cm}^{-3}$ is consistent with the
typical density of $\gtrsim$$10^9\,\mathrm{cm}^{-3}$ expected for BLR
clouds \citep{Baldwin2003,Osterbrock1989}.

\subsection{The morphology and spatial distribution of a putative
  cloudy absorber} 
\label{sec:clumpytorus}

We now investigate the possibility for the detected absorber to be
part of an overall clumpy medium. 

According to the clumpy torus model of
\citet{Nenkova2008a,Nenkova2008b}, the average number clouds along a
line of sight with inclination $i$ with respect to the system symmetry
axis, $N_\mathrm{C}$, is given by
\begin{equation}
  \label{eq:num_clouds}
  N_\mathrm{C}=N_0
  \exp{\left[-\left(\frac{90-i}{\sigma}\right)^{2}\right]}
\end{equation}
where $N_0$ is the number of clouds along a ray in the equatorial
plane between the $R_\mathrm{d}$ and $Y\,R_\mathrm{d}$. The radial
distribution of the clouds follows a power law $r^{-q}$. This model
has observational support from extensive time-resolved X-ray
spectroscopy \citep{Markowitz2014} and Bayesian model fits
\citep{AsensioRamos2009} to IR SEDs
\citep{RamosAlmeida2011,AlonsoHerrero2011}.

\citet{Elitzur2007} and \citet{Gaskell2008} claim that clouds can
exist both in the BLR below, and in the dusty torus beyond
$R_\mathrm{d}$, all as part of a common structure. We therefore assume
for simplicity that the power-law index $q$ is the same for both the
BLR and the torus clouds, which yields a successively growing number
of clouds towards the center. We assume the following mode values for
the clumpy torus parameters from the posterior probability
distributions of \citet{AlonsoHerrero2011}: $N_{0}=15$, $Y=15$,
$q=0.1$, $\sigma=44\degr$ and $i=30\degr$. They use IR photometry and
additionally spectroscopic data around 10\,$\mu\mathrm{m}$. These
values result in $\sim$16 clouds in the equatorial plane between 0.4
and $15R_\mathrm{d}$. The lower limit equals the derived minimal
distance of the absorber in NGC~3227. We call this number
$N_{0}^\mathrm{X+IR}$, as it includes both dusty clouds detected in
the IR and also dust-free clouds that additionally absorb the
X-rays. The number of clouds at the given inclination angle and width
of the Gaussian cloud distribution then equals
$N_\mathrm{C}^\mathrm{X+IR}=2.5$ (Eq.~12).

For comparison one can estimate the average number of clouds on the
line of sight based on 6.9\,years of sustained \rxte monitoring, the
2006 \xmm observation, and the 35\,days of \suzaku and \swift\
monitoring in 2008. During that time NGC~3227 was in an obscured state
for a total of at least $\sim$114\,days. Based on the given data,
NGC~3227 spends 4.4\% of the observed time in eclipse.

Assuming Poisson statistics, the time averaged escape probability for
photons without undergoing strong absorption by an average of
$N_\mathrm{C}^\mathrm{X+IR}$ clouds in the line of sight is given by
\begin{equation}
  \label{eq:escapeprobab}
  P\sim \exp\left(-N_\mathrm{C}^\mathrm{X+IR}\,\tau_\lambda\right)
\end{equation}
where the monochromatic optical depth of one cloud,
$\tau_\lambda\gtrsim 1$ in order to be able to obscure the line of
sight as observed during eclipses. With a 4.4\% probability of
obscuration, the integrated escape probability equals $\sim$95.5\%,
i.e., $N_\mathrm{C}^\mathrm{X+IR}\sim 0.045$. This value is clearly
lower than the average of 2.5\,clouds estimated above. 

One can still find consistency when considering severe degeneracies
between different attempts of clumpy tori
fits. \citet{RamosAlmeida2011} for example only use IR photometry
between 1--$40\,\mu\mathrm{m}$. They find much larger uncertainties
and different mode values of $N_{0}=2$, $Y=19$, $q=0.6$,
$\sigma=33\degr$ and $i=66\degr$. Based on these values we find
$N_{0}^\mathrm{X+IR}=2.3$ between 0.4 to $30\,R_\mathrm{d}$;
$N_\mathrm{C}^\mathrm{X+IR}$ is consequently equal to 1.4. When
inserting not the mode values but other values of high probability
within the posterior probability distributions of the parameters, we
can find consistency with the observed number of 0.045 clouds along
the line of sight. If the inclination angle is even more face-on, the
number of clouds $N_\mathrm{C}^\mathrm{X+IR}$ on the line of sight is
also significantly reduced. \citet{Marin2014} provides an overview
over different methods and inclination angles measured so
far. \citet{Hicks2008} and \citet{Fischer2013} indeed find a value of $i\sim 16\degr$ using
near-IR spectroscopy consistent with \citet{Fischer2013} although the
results from the optical polarimetry (Sect.~\ref{sec:opt_pol}) are
more consistent with the intermediate inclinations derived from IR SED
fitting.

We note that the fact that we obtain a larger number of clouds in the
line of sight from extending the clumpy torus model than required from
historical eclipse events is not inconsistent with the spectral
results. \citet{Markowitz2009} show that even in a relatively
unabsorbed state, ionized absorbers are present in the line of sight.
Those could explain the determined excess of clouds.

\subsection{The origin of the absorber}
Here, we discuss the implications of the measured variable density
profile of the absorber. Besides the strongly variable column density
over time, the covering fraction remains roughly constant for the
latter five out of six \suzaku observations. This argues against two
distinct clouds as explanation for the two overdensities but for a
potential filament with a variable internal density structure that
enters the line of sight. As a test, we can estimate the limiting size
of one cloud, which is confined by its own magnetic field or the
external pressure of the intercloud medium, not to get tidally sheared
by the central force of gravity, to
\begin{equation}
  \label{eq:size_shear}
  r_\mathrm{cl}\leq 10^{16}\frac{N_{23} R_\mathrm{pc}^{3}}{M_{7}}\,\mathrm{cm}
\end{equation}
\citep{Elitzur2006} where
$N_{23}=N_\mathrm{H}/10^{22}\,\mathrm{cm}^{-2}$,
$R_\mathrm{pc}=R/1\,\mathrm{pc}$ is the distance of the cloud to the
central engine, and where $M_{7}=M_\mathrm{BH}/10^{7}\,M_{\astrosun}$.
Inserting the average column density $N_\mathrm{H}$ from the 2008
event and $R_\mathrm{pc}\sim 0.017$--0.65, we find $r_\mathrm{cl}\leq
3.3\times 10^{10}-1.8\times 10^{15}\,\mathrm{cm}$. For comparison, the
estimated diameter of a spherical cloud on a Keplerian orbit was
$6.6\times 10^{13}\,\mathrm{cm}$--$4.1\times 10^{14}\,\mathrm{cm}$.
Both ranges are consistent. Note, however, that the limiting size of
the absorber, $r_\mathrm{cl}$, is strongly dependent on its distance.
We can argue for a distance $R_\mathrm{pc}$ of the absorbing cloud
towards the lower limit of allowed ranges when considering the weak
reddened, implying a location inside the dust sublimation radius.  In
this case we would have to assume a rather extended and filamentary
medium in contrast to the simplified picture of spherically symmetric
clouds bound by self-gravity.

Mechanisms that are thought to be able to lead to such structures
include MHD-driven winds that are launched from the accretion disk,
for example, via magneto-centrifugally acceleration
\citep{BP1982,Contopoulos1994,Konigl1994}. We find consistency between
the measured ionization structure and column density of the absorber
and the range of values simulated for MHD winds by
\citet{Fukumura2010}.

The absorber is also qualitatively consistent
with the turbulent, dusty disk wind proposed by \citet{Czerny2011}. In
this model, dusty clouds from the low-ionization part of the BLR can
rise from the disk, e.g., driven by MHD-forces, and get irradiated
and heated sufficiently such that dust sublimates. Eventually clouds
may fall back towards the disk when the gravitational force prevails
over the driving forces. In this context the absorbing cloud from 2008
may be just recently sublimated and temporarily situated at a height
intercepting our line of sight.

The question remains how such driven clouds can get stretched out to
filamentary structures of internal overdensities that seem to best
explain the observations. A potential mechanism is described by
\citet{Emmering1992} in the context of MHD winds. As a wind or cloud
forms, the material initially contains a high amount of dust, which is
a good coolant via IR thermal emission. Once the clouds are driven
away from the accretion disk they are heating up to
$\sim$$10^{4}\,\mathrm{K}$ as the dust sublimes. After a cloud is
photo-ionized, its internal pressure increases and expands at roughly
its internal sound speed until its pressure decreases to roughly that
exerted by the external magnetic field. \citet{Rees1987} show that the
polodial magnetic field of the disk together with the internal field
of magnetized BLR clouds can cause the expanded clouds to stretch out
along the magnetic field, which is frozen in the disk.

Additionally we mention radiative driving of accretion disk winds.
\citet{Dorodnitsyn2012} provide detailed simulations on how the X-ray
radiation from the hot corona can be transferred to the IR by exerting
pressure on dust and allowing clumpy or filamentary structures to
arise. Again, the cloud or filament can be ionized by the nuclear
radiation when reaching a certain height matching the observables of
the absorber.

The 2008 absorption event might not be unique. Similar to the
long-term obscurations of $\sim$80\,d and $\geq 35\,\mathrm{d}$ of
NGC~3227 in 2000/2001 and 2008, respectively, NGC~5548 is continuously
obscured for a few years \citep{Kaastra2014}. Such phenomena may thus
be typical to Seyfert galaxies although the probability to observe
them is low.

\section{Summary and Conclusions}
We have studied data from a 5\,week long \suzaku and \swift\ X-ray
and UV monitoring campaign on the Sy 1.5 NGC~3227 in late 2008, which
caught the source in an absorbed state. We performed a time-resolved
X-ray spectroscopy to untangle the various emission components, i.e.,
the coronal power law and ionized reflection, as well as multiple ionized
absorption components, and isolate the time-dependent behavior of the
dominant absorber in 2008. Past \rxte monitoring reveals NGC~3227 to
be usually unabsorbed, except for eclipses in 2000/2001
\citep{Lamer2003} and 2002 \citep{Markowitz2014}.  The ultimate aim of
this study is to understand the properties of the X-ray absorber.

We used two archival and previously published \xmm spectra to derive
a baseline model that we fit simultaneously to six \suzaku and twelve
\swift\ observations from 2008. An extensive exploration of the
parameter space reveals only minor degeneracies. The two \xmm spectra
allowed us to identify three differently ionized absorbers, called
\WA{1} (lowly ionized), \WA{2} (intermediately ionized) and \WA{3}
(highly ionized). We conclude that the detected absorption variability
in 2008 is solely due to \WA{2}, while WA1 and \WA{3} remain
constant. 

We resolve the column density and covering fraction of \WA{2} to be
variable with time. The covering fraction rises from $\sim$70\% for
the first \suzaku observation up to $\sim$90\% for the remaining five
observations. The column density varies from $\sim$5 to
$\sim$$18\times 10^{22}\,\mathrm{cm}^{-1}$ in a doubly-peaked density
profile. The ionizing luminosity $L_\mathrm{ion}$ and the ionization
state $\xi$ of \WA{2} are seen to vary by factors of 1.3 and 104,
respectively. A correlation analysis shows a tentative dependence
between $\xi$ and $L_\mathrm{ion}$ as a proxy for $L_\mathrm{ion}$.

We use the average parameters of \WA{2} ($\langle
N_\mathrm{H}\rangle\sim 11.2\times 10^{22}\,\mathrm{cm}^{-2}$,
$\langle\log \xi\rangle\sim 1.1$, $\langle L_\mathrm{ion}\rangle \sim
8.9\times 10^{42}\,\mathrm{erg}\,\mathrm{s}^{-1}$) during the
$\geq$35\,d long absorption event to estimate that the distance of the
absorbing medium is between 0.017 and 0.65\,pc from the central SMBH.
Here we made use of all available information, which comprise the
ionization state, column density and minimal duration of the event as
well as the incident irradiation. The underlying, simplified
assumption is that of a spherical cloud on a Keplerian orbit around
the illuminating source.

For the derived distance range we infer the hydrogen number density of
the spherical cloud to be $\sim 2.7\times 10^{8}-1.7\times
10^{9}\,\mathrm{cm}^{-3}$. The cloud therefore has a diameter of
$\Delta R \sim 6.6\times 10^{13}-4.1\times 10^{14}\,\mathrm{cm}$. We
conclude that the absorber may be located in the outermost dust free
BLR or the inner dusty torus.

We find only moderate reddening in the \swift\ UVOT data, which is due
to a column density of $N_\mathrm{H}\sim 1.2\times
10^{21}\,\mathrm{cm}^{-2}$. In contrast, we measure X-ray-absorbing
column densities being about a factor of 100 larger. We conclude based
on Sect.~\ref{sec:origin} that the X-ray absorber responsible for the
variability in 2008 and the absorber responsible for extinction are
distinct. The first seems to comprise a high gas-to-dust ratio due to
its vicinity to the SMBH. A location in the BLR would be consistent with
our results. The latter is consistent with the distant (100\,pc)
lukewarm absorber proposed by \citet{Kraemer2000} and
\citet{Crenshaw2001}, which still seems to be present in recent
\swift\ observations.

We also investigated the measured absorber in the context of the
clumpy absorber model of \citet{Nenkova2008a,Nenkova2008b}. The
inferred distribution of clouds is consistent with the eclipses
observed for NGC~3227. The result that the absorber is likely located
in the BLR, leads us to extend the distribution of clouds also down to
$0.4R_\mathrm{d}$.  The consistency of the model with past
observations and the 2008 event is in particular reached for small
inclination angles ($\lesssim20$--$30\degr$) of the putative clumpy
torus that are measured by independent methods and help to explain the
predominant lack of absorption in NGC~3227 during past observations.

A spherical cloud situated below the dust sublimation zone can most
likely not withstand tidal shearing by the central source of gravity.
Several mechanisms including MHD-driving are able to lift an initially
dusty cloud from the disk where it is getting exposed to nuclear
radiation and therefore ionized. Pressure gradients potentially
stretch the cloud preferably along the magnetic field lines frozen to
the disk.

In that context, the observing campaign represents a series of
snapshots of parts of a cloud or filament whose properties (size,
density distribution) may be evolving over time. The density profile
of this event contrasts with that of the 2000/2001 event, which showed
a centrally symmetric profile and clear ingress and egress. We
therefore conclude that NGC~3227 is a rare laboratory to study the
range of physical processes that form and shape clouds or filaments.

\begin{acknowledgements}
  This research was partially funded by a grant from the
  Bundesministerium f\"ur Wirtschaft und Technologie under Deutsches
  Zentrum f\"ur Luft- und Raumfahrt grant 50\,OR\,1311. We thank Mirko
  Krumpe and Robert Nikutta for reading the manuscript and giving very
  helpful comments. We made use of ISIS functions provided by
  ECAP/Remeis observatory and MIT
  (\url{http://www.sternwarte.uni-erlangen.de/isis/}) as well as the
  NASA/IPAC Extragalactic Database (NED) which is operated by the Jet
  Propulsion Laboratory, California Institute of Technology, under
  contract with the National Aeronautics and Space Administration.
  This work used data obtained with the \suzaku satellite, a
  collaborative mission between the space agencies of Japan (JAXA) and
  the USA (NASA), as well as data from the \swift\ satellite, a NASA
  mission managed by the Goddard Space Flight Center, \xmm, an ESA
  science mission with instruments and contributions directly funded
  by ESA Member States and NASA, the Wide-field Infrared Survey
  Explorer, which is a joint project of the University of California,
  Los Angeles, and the Jet Propulsion Laboratory/California Institute
  of Technology, funded by NASA, and the Two Micron All Sky Survey, a
  joint project of the University of Massachusetts and the Infrared
  Processing and Analysis Center/California Institute of Technology,
  funded by NASA and the National Science Foundation. We also used
  observations made with the NASA/ESA Hubble Space Telescope, obtained
  from the data archive at the Space Telescope Science Institute.
  STScI is operated by the Association of Universities for Research in
  Astronomy, Inc.\ under NASA contract NAS 5-26555. The research is
  also based upon observations acquired at the Observatorio
  Astron\'omico Nacional in the Sierra San Pedro M\'artir (OAN-SPM),
  Baja California, M\'exico. TM acknowledges support by UNAM-DGAPA
  Grant PAPIIT IN104113 and CONACyT Grant Cient\'ifica B\'asica
  179662. We thank J. E. Davis for the development of the
  \texttt{slxfig} module that has been used to prepare the figures in
  this work.
\end{acknowledgements}

\bibliographystyle{jwaabib}

\newpage
\onecolumn
\begin{landscape}
\section{Appendix}
\label{sec:appendix}

\begin{table}
  \caption{List of parameters for the simultaneous fit of observations
    Suz~2 to Suz~6 and the individual fit of the observation Suz~1. We
    use the same model for both fits. The simultaneous fit consists of
    time-dependent parameters that are equal for a group comprising
    XIS and HXD data of one observation and time-independent parameters that are
    equal for all data, denoted with quotation marks. Frozen
    parameters are denoted by the symbol \textdagger, $\ast$ denotes
    parameters adopted from the simultaneous fit. The redshift is
    frozen to the systemic value of $z=0.003859$
    \citep{Vaucouleurs1991}. All spectra are additionally absorbed by
    a cold column of Galactic absorption of $1.99\times
    10^{20}\,\mathrm{cm}^{-2}$ \citep{Kaberla2005}. Normalizations are
    given in units of
    $\mathrm{ph}\,\mathrm{keV}^{-1}\,\mathrm{cm}^{-2}\,\mathrm{s}^{-1}$
    at 1\,keV.}\label{tab:par_suz} \def\arraystretch{1.2}
\begin{tabular}{lllllllll}
\hline\hline
\multicolumn{9}{c}{\texttt{detconst(1)$\ast$(cutoffpl(\%)+xillver(\%))$\ast$((1-constant(\%))+constant(\%)$\ast$zxipcfTB(2+\%))$\ast$zxipcfTB(1)$\ast$zxipcfTB(2)$\ast$tbnew\_simple(1)}}\\
\hline
& & Suz~1 & Suz~2 & Suz~3 & Suz~4 & Suz~5 & Suz~6 & \\
\hline
PL & norm ($\times 10^{-2}$) & $1.01\pm0.04$& $0.62^{+0.04}_{-0.06}$& $0.76^{+0.05}_{-0.04}$& $0.345^{+0.046}_{-0.026}$& $0.82^{+0.09}_{-0.05}$& $0.70^{+0.11}_{-0.08}$& \parbox[c]{2mm}{\multirow{7}{*}{\rotatebox[origin=c]{-90}{time dependent}}}\\
     & $\Gamma$& $1.658\pm0.017$& $1.592^{+0.026}_{-0.036}$& $1.653^{+0.031}_{-0.025}$& $1.62^{+0.05}_{-0.04}$& $1.694^{+0.041}_{-0.029}$& $1.72^{+0.06}_{-0.05}$& \\
Ion.\ Refl. & norm ($\times 10^{-5}$) & $0.31^{+0.51}_{-0.11}$& $0.24\pm0.05$& $0.192^{+0.179}_{-0.024}$& $0.29^{+0.18}_{-0.05}$& $0.37^{+0.43}_{-0.13}$& $0.60^{+15.5}_{-0.40}$& \\
           & $\log\xi$\,($\mathrm{erg}\,\mathrm{cm}\,\mathrm{s}^{-1}$)& $1.73^{+0.19}_{-0.44}$& $2.05^{+0.09}_{-0.05}$& $2.03^{+0.04}_{-0.29}$& $2.01^{+0.04}_{-0.25}$& $1.69^{+0.17}_{-0.37}$& $1.4^{+0.4}_{-1.4}$& \\
\WA{2} & $N_\mathrm{H}$\,($10^{22}\mathrm{cm}^{-2}$)& $4.75^{+0.32}_{-0.29}$& $16.0^{+1.0}_{-0.8}$& $6.2\pm0.5$& $15.9^{+1.6}_{-1.5}$& $10.8\pm0.7$& $13.4^{+1.0}_{-0.9}$& \\
 & $\log\xi$\,($\mathrm{erg}\,\mathrm{cm}\,\mathrm{s}^{-1}$)& $2.04\pm0.04$& $1.27^{+0.12}_{-0.08}$& $0.71^{+0.29}_{-0.23}$& $0.08^{+0.19}_{-0.17}$& $1.21\pm0.04$& $1.11^{+0.07}_{-0.16}$& \\
 & $f_\mathrm{cvr}$& $0.735^{+0.024}_{-0.025}$& $0.873^{+0.007}_{-0.012}$& $0.905\pm0.006$& $0.833^{+0.016}_{-0.013}$& $0.929\pm0.007$& $0.930\pm0.009$& \\
\hline
Detconst & XIS\,0 & \multicolumn{1}{l}{$1^{\dagger}$} & $1^{\dagger}$ & \textquotedbl & \textquotedbl & \textquotedbl & \textquotedbl & \parbox[c]{2mm}{\multirow{13}{*}{\rotatebox[origin=c]{-90}{time-independent}}}\\
 & XIS\,1 & \multicolumn{1}{l}{$1.058^{+0.010}_{-0.008}$} & $1.096\pm0.007$ & \textquotedbl & \textquotedbl & \textquotedbl & \textquotedbl &\\
 & XIS\,3 & \multicolumn{1}{l}{$1.054^{+0.009}_{-0.011}$} & $1.036\pm0.006$ & \textquotedbl & \textquotedbl & \textquotedbl & \textquotedbl &\\
 & HXD & \multicolumn{1}{l}{1.16$^{\dagger}$} & $1.16^{\dagger}$ & \textquotedbl & \textquotedbl & \textquotedbl & \textquotedbl &\\
Ion.\ Refl. & $Z_\mathrm{Fe}$ & \multicolumn{1}{l}{$2.81^{\ast}$} & $2.81\pm0.17$ & \textquotedbl & \textquotedbl & \textquotedbl & \textquotedbl &\\
\WA{1} & $N_\mathrm{H}$\,($10^{22}\mathrm{cm}^{-2}$) & \multicolumn{1}{l}{$0.137^{+0.009}_{-0.006}$} & $0.068^{+0.025}_{-0.014}$ & \textquotedbl & \textquotedbl & \textquotedbl & \textquotedbl &\\
 & $\log\xi$\,($\mathrm{erg}\,\mathrm{cm}\,\mathrm{s}^{-1}$) & \multicolumn{1}{l}{$-0.29^{+0.09}_{-0.13}$} & $-0.9\pm0.6$ & \textquotedbl & \textquotedbl & \textquotedbl & \textquotedbl &\\
 & $f_\mathrm{cvr}$ & \multicolumn{1}{l}{$1.00^{\dagger}$} & $1.00^{\dagger}$ & \textquotedbl & \textquotedbl & \textquotedbl & \textquotedbl &\\
\WA{3} & $N_\mathrm{H}$\,($10^{22}\mathrm{cm}^{-2}$) & \multicolumn{1}{l}{$3.0^{+0.8}_{-1.3}$} & $4.1^{+4.1}_{-2.1}$ & \textquotedbl & \textquotedbl & \textquotedbl & \textquotedbl &\\
 & $\log\xi$\,($\mathrm{erg}\,\mathrm{cm}\,\mathrm{s}^{-1}$) & \multicolumn{1}{l}{$3.44^{+0.07}_{-0.05}$} & $4.17^{+0.16}_{-0.22}$ & \textquotedbl & \textquotedbl & \textquotedbl & \textquotedbl &\\
 & $f_\mathrm{cvr}$ & \multicolumn{1}{l}{$1.00^{\dagger}$} & $1.00^{\dagger}$ & \textquotedbl & \textquotedbl & \textquotedbl & \textquotedbl &\\
\hline
\multicolumn{2}{l}{$L^\mathrm{ion}_{0.0136-13.6\,\mathrm{keV}}\,(10^{42}\mathrm{erg}\,\mathrm{s}^{-1})$} & $7.6\pm 0.03$ & $7.4\pm 0.05$ & $7.3\pm 0.04$ & $5.6\pm 0.05$ & $6.7\pm 0.04$ & $5.6\pm 0.04$ & \\
\multicolumn{2}{l}{$L^\mathrm{X}_{0.6-10\,\mathrm{keV}}\,(10^{42}\mathrm{erg}\,\mathrm{s}^{-1})$} & $2.60\pm 0.03$ & $1.80\pm 0.05$ & $2.09\pm 0.04$ & $9.82\pm 0.05$ & $2.15\pm 0.04$ & $1.80\pm 0.04$ & \\
\end{tabular}
\end{table}

\begin{table}
  \caption{List of parameters for the simultaneous fit of all
    \swift\ observations. The model equals the one fitted to the
    \suzaku data. The simultaneous fit consists of time-dependent
    parameters individually fitted to each observation and time-independent
    parameters that are equal for all observations, denoted with
    quotation marks. The symbol $\ast$ marks parameters adopted from
    the simultaneous fit of the \suzaku observations. The redshift is
    frozen to the systemic value of $z=0.003859$ \citep{Vaucouleurs1991}.
    All spectra are additionally absorbed by a cold column of Galactic
    absorption of $1.99\times 10^{20}\,\mathrm{cm}^{-2}$
    \citep{Kaberla2005}. Normalizations are
    given in units of
    $\mathrm{ph}\,\mathrm{keV}^{-1}\,\mathrm{cm}^{-2}\,\mathrm{s}^{-1}$
    at 1\,keV.}\label{tab:par_swift}
\centering
\renewcommand{\arraystretch}{1.3}
\begin{tabular}[ht]{llllllll}
\hline\hline
& & Sw~1a & Sw~2a & Sw~2b & Sw~3a & Sw~3b & \\
\hline
PL & norm & $\left(0.84^{+0.14}_{-0.20}\right)\times10^{-2}$& $0.008^{+0.010}_{-0.004}$& $\left(3.5^{+1.5}_{-1.7}\right)\times10^{-3}$& $\left(0.78^{+0.16}_{-0.20}\right)\times10^{-2}$& $\left(0.69^{+0.10}_{-0.11}\right)\times10^{-2}$& \multirow{4}{*}{\rotatebox[origin=c]{-90}{time dep.}}\\
Ion.\ Refl. & norm& $\left(0.28^{+0.93}_{-0.28}\right)\times10^{-5}$& $\left(0.5^{+1.3}_{-0.5}\right)\times10^{-5}$& $\left(0.9^{+0.6}_{-0.7}\right)\times10^{-5}$& $\left(0.24^{+1.02}_{-0.25}\right)\times10^{-5}$& $\le0.6\times10^{-5}$& \\
\WA{2} & $N_\mathrm{H}$\,($10^{22}\mathrm{cm}^{-2}$)& $6.8^{+6.6}_{-2.8}$& $27^{+25}_{-8}$& $15^{+8}_{-11}$& $8.3^{+4.3}_{-2.9}$& $7.0^{+2.4}_{-1.5}$& \\
 & $f_\mathrm{cvr}$& $0.80^{+0.14}_{-0.09}$& $0.90^{+0.06}_{-0.07}$& $0.79^{+0.05}_{-0.13}$& $0.86^{+0.08}_{-0.06}$& $0.93^{+0.05}_{-0.04}$& \\
\hline
& & Sw~4a & Sw~4b & Sw~5a & Sw~5b & Sw~6b\\
\hline
PL & norm& $\left(0.36^{+0.40}_{-0.25}\right)\times10^{-2}$& $0.007^{+0.011}_{-0.006}$& $\left(0.62^{+0.12}_{-0.21}\right)\times10^{-2}$& $\left(0.39^{+0.13}_{-0.17}\right)\times10^{-2}$& $\left(3.6^{+1.4}_{-0.9}\right)\times10^{-3}$& \multirow{4}{*}{\rotatebox[origin=r]{-90}{time dep.}}\\
Ion.\ Refl. & norm& $\left(1.1^{+0.7}_{-0.9}\right)\times10^{-5}$& $\le1.2\times10^{-5}$& $\left(0.19^{+0.68}_{-0.19}\right)\times10^{-5}$& $\left(0.017^{+0.533}_{-0.017}\right)\times10^{-5}$& $\left(0.26^{+0.54}_{-0.27}\right)\times10^{-5}$& \\
\WA{2} & $N_\mathrm{H}$\,($10^{22}\mathrm{cm}^{-2}$)& $42^{+22}_{-30}$& $54^{+45}_{-38}$& $15^{+4}_{-6}$& $15^{+6}_{-8}$& $8.2^{+5.5}_{-2.5}$& \\
 & $f_\mathrm{cvr}$& $0.85^{+0.08}_{-0.25}$& $0.903^{+0.015}_{-0.404}$& $0.911^{+0.023}_{-0.037}$& $0.90^{+0.04}_{-0.07}$& $0.93^{+0.07}_{-0.04}$& \\
\midrule
PL & $\Gamma$ & $1.50\pm0.04$ & \textquotedbl & \textquotedbl & \textquotedbl & \textquotedbl & \multirow{10}{*}{\rotatebox[origin=c]{-90}{time-independent}}\\
Ion. Refl. & $\log\xi$\,($\mathrm{erg}\,\mathrm{cm}\,\mathrm{s}^{-1}$) & $1.84\pm0.15$ & \textquotedbl & \textquotedbl & \textquotedbl & \textquotedbl &\\
           & $Z_\mathrm{Fe}$ & $2.81^{\ast}$ & \textquotedbl & \textquotedbl & \textquotedbl & \textquotedbl &\\
\WA{2} & $\log\xi$\,($\mathrm{erg}\,\mathrm{cm}\,\mathrm{s}^{-1}$) & $1.44\pm0.12$ & \textquotedbl & \textquotedbl & \textquotedbl & \textquotedbl &\\
\WA{1} & $N_\mathrm{H}$\,($10^{22}\mathrm{cm}^{-2}$) & $0.066^{+0.018}_{-0.017}$ & \textquotedbl & \textquotedbl & \textquotedbl & \textquotedbl &\\
     & $\log\xi$\,($\mathrm{erg}\,\mathrm{cm}\,\mathrm{s}^{-1}$) & $-0.89^{\ast}$ & \textquotedbl & \textquotedbl & \textquotedbl & \textquotedbl &\\
     & $f_\mathrm{cvr}$ & 1$^{\ast}$ & \textquotedbl & \textquotedbl & \textquotedbl & \textquotedbl &\\
\WA{3} & $N_\mathrm{H}$\,($10^{22}\mathrm{cm}^{-2}$) & $4.12^{\ast}$ & \textquotedbl & \textquotedbl & \textquotedbl & \textquotedbl &\\
     & $\log\xi$\,($\mathrm{erg}\,\mathrm{cm}\,\mathrm{s}^{-1}$) & $4.17^{\ast}$ & \textquotedbl & \textquotedbl & \textquotedbl & \textquotedbl &\\
     & $f_\mathrm{cvr}$ & $1^{\ast}$ & \textquotedbl & \textquotedbl & \textquotedbl & \textquotedbl &\\
\hline
\end{tabular}
\end{table}
\end{landscape}
\begin{figure*}
  \centering
  \resizebox{\hsize}{!}{\includegraphics{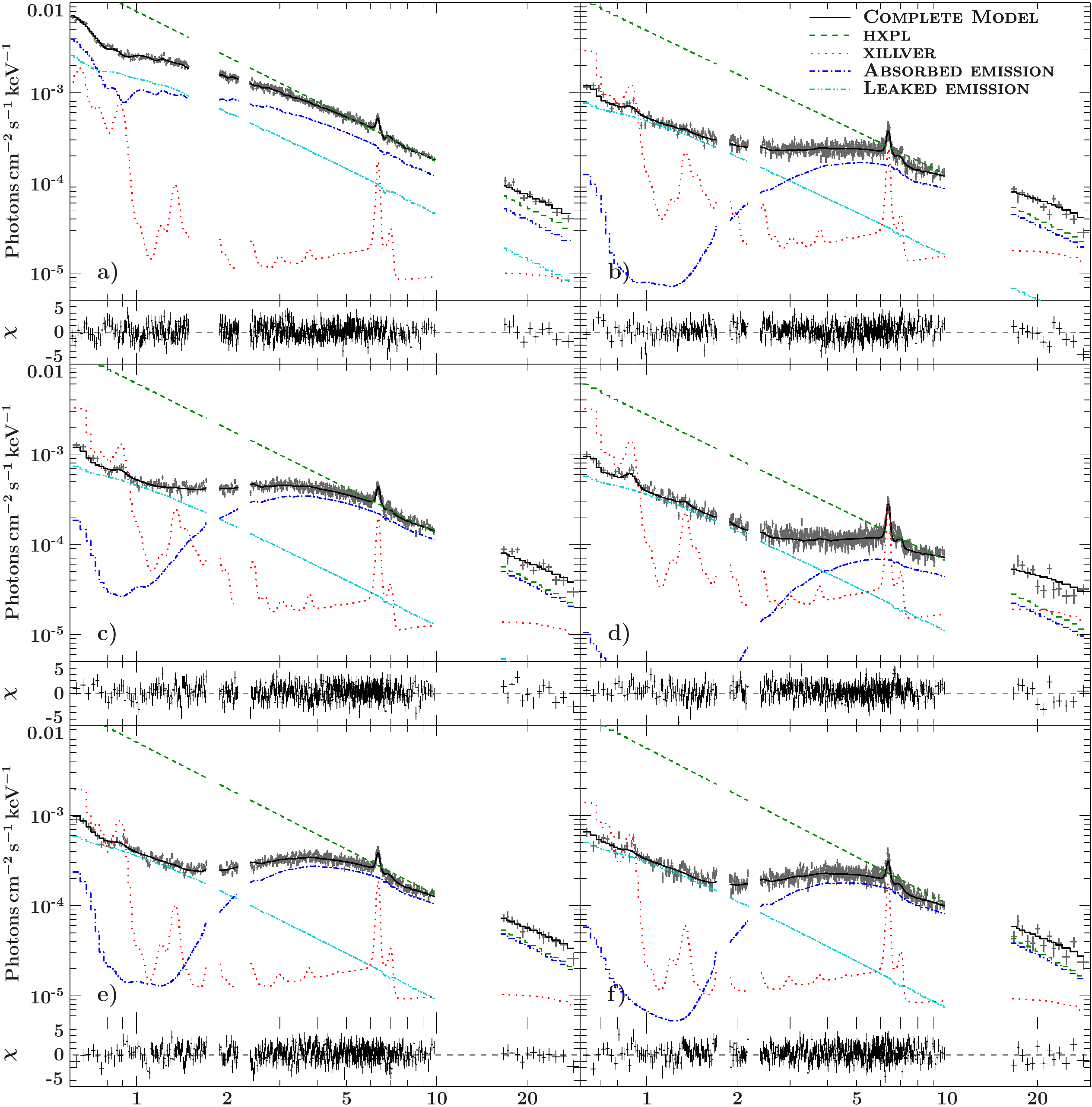}}
  \caption{Spectra and model components of all \suzaku observations.
    We show the complete model (black solid line) with residuals, the
    hard X-ray power law (dark green dashed line), the ionized
    reflection component (red dotted line), the emission absorbed by
    the variable partially covering column (blue dotted-dashed line)
    and the leaked emission (cyan doubly-dotted-dashed line).}
  \label{fig:all_suz_comp}
\end{figure*}

\end{document}